\documentclass[11pt,a4paper]{article}

\usepackage[utf8]{inputenc}
\usepackage{newtxtext,newtxmath}
\usepackage{graphicx}
\usepackage[margin=1in]{geometry}
\usepackage[labelfont=bf]{caption} 
\usepackage{cite} 
\usepackage{url}
\usepackage{authblk} 
\title{\textbf{Fluctuating growth rate and spatial diffusion shape plankton diversity}}

\author[1]{Giorgio Vittorio Visco}
\author[1]{Kobe Simoens}
\author[2]{Emanuele Pigani}
\author[3]{Diana Sarno}
\author[1,4]{Samir Suweis}
\author[3,*]{Daniele Iudicone}
\author[1,4,\dag]{Sandro Azaele}

\affil[1]{\small Dipartimento di Fisica e Astronomia "Galileo Galilei", University of Padua, Padua, Italy.}
\affil[2]{\small Quantitative Life Sciences, The Abdus Salam International Centre for Theoretical Physics, Trieste, Italy.}
\affil[3]{\small Stazione Zoologica Anton Dohrn, Naples, Italy.}
\affil[4]{\small INFN Sezione di Padova, Padua, Italy.}
\affil[*]{\small Corresponding author. Email: daniele.iudicone@szn.it}
\affil[$\dagger$]{\small Corresponding author. Email: sandro.azaele@unipd.it}

\date{}

\begin{document}

\maketitle

\begin{abstract} \bfseries \boldmath

Planktonic communities exhibit ubiquitous population distributions and patchy spatial  structures, yet the fundamental mechanisms driving them remain debated. Here, we derive these regularities from a minimalistic theoretical description that incorporates stochastic fluctuations in growth rates and effective ocean dispersal. We combine global metabarcoding, microscopy, and high-resolution chlorophyll datasets and show that the decay of spatial correlations, the crossover regimes of Taylor's law, the patterns of local species diversity and biomass distributions agree with common underlying dynamics. These results suggest that the intertwined effect of diffusivity and fluctuating growth rate, captured by an emergent correlation length, shapes plankton spatial heterogeneity from local to long-range scales, reconciling local variability with macroecological patterns.
\end{abstract}

\noindent
Plankton communities play a critical role in marine ecosystems as they drive biogeochemical cycles~\cite{falkowski1998_biogeo, siegel2023_quantifying}, sustain higher trophic levels~\cite{tagliabue2021_persitent}, and serve as sensitive indicators of ocean health~\cite{hays2005_climate, ratnarajah2023_monitoring}. As such, the quantitative modeling of planktonic systems has become essential not only for advancing the mechanistic understanding of marine biodiversity~\cite{follows2007_emergent, ward2012_size} but also for informing its conservation and management~\cite{stock2011_ipcc}. Nevertheless, the characterization of such systems is constrained by the complexity of the seascape~\cite{steele1989_ocean}, i.e., the highly heterogeneous environment across space and time where biological processes and fluid physics jointly shape the ecology of plankton. Fluctuations in environmental factors, including temperature, salinity, and resource abundance, contribute to the spatiotemporal variability that is reflected in the fitness of different species of plankton~\cite{mustonen2009_fitness}. At the same time, plankton community structure is shaped by multiple biotic interactions, for example, viral infection~\cite{suttle2007_virus}, which tends to disproportionately affect fast-growing taxa; and top-down predation~\cite{verity1996_organism}, which regulates trophic dynamics and ecosystem organization.

Considerable effort has been devoted to identifying the patterns that best quantify plankton biodiversity and its main drivers. A common measure of biodiversity is the imbalance between hyperdominant species and species with very low population sizes~\cite{vannes2024_tiny}. Generally, ecosystems are characterized by uneven distributions where most organisms belong to a few species, and most of the species are very rare~\cite{gao2025_powerbend}. This pattern of unevenness is typically quantified by the Species Abundance Distribution (SAD)~\cite{McGill2007_species}, i.e., the full distribution of relative abundances among species.  In this context, Neutral Theory~\cite{Volkov2003_neutral, Azaele2016_statmech,behrenfeld2024_2024} has proven to be a valuable tool for describing and interpreting SADs and many other biodiversity patterns~\cite{Azaele2016_statmech}. In neutral models, the abundances of species are viewed as outcomes of stochastic birth-and-death processes that do not take into account the species' identity of individual organisms. Although it is not mandatory from the neutral assumption, often these models assume that the variability of abundance among species is essentially due to demographic stochasticity, which originates from the intrinsic stochastic nature of birth/death events and speciation/immigration. In recent years, this framework has been successfully applied to a wide range of ecological systems, including planktonic communities~\cite{SerGiacomi2018_SAD, Pigani2024_diatoms_neutral}. However, this approach also has some limitations, as it does not account for mixing effects of oceanic currents. Recent studies~\cite{villa2020_ocean} have illustrated the possible consequences of turbulent regimes on community diversity, but analytical tools to quantify them remain lacking. Another important drawback is that classical demographic-stochasticity models typically assume that birth and death rates are constant over time, so that the net growth rate does not fluctuate~\cite{Azaele2016_statmech}. However, this assumption is not consistent with temporal fluctuations commonly observed in ecosystems~\cite{Kalyuzhny2015_neutral_theory, vannes2024_tiny, danino2016_effect}, motivating an alternative in which the growth rate itself is allowed to vary--the
so-called \emph{environmental noise}~\cite{Dennis1991_estimation,  Lande2003_Stochastic}. 

The relevance of this framework to planktonic systems is motivated by the empirical observation that phytoplankton biomass time series are typically log-normally distributed~\cite{Campbell1995_lognormal}, a pattern consistent with Gaussian fluctuations in growth rate. Although environmental noise has been used to describe temporal patterns such as turnover rates~\cite{mallmin2024_chaotic}, its role in shaping spatial heterogeneity remains much less understood. Recently, some studies~\cite{swartz2022_seascape, amer2026_spatiotemporal} have combined stochastic growth rates with spatial diffusion to characterize plankton communities, while classical studies of plankton patchiness have shown that large-scale spatial structure emerges from the interplay between population dynamics and ocean transport~\cite{abraham1998_generation,mahadevan2002_biogeochemical,Bracco2009_horizontal}. Yet these two perspectives have remained largely disconnected: there is still no general analytical framework that links local growth-rate fluctuations, spatial dispersal, and the emergent macroecological statistics measured across marine seascapes. In particular, it remains unclear whether spatial correlations, abundance distributions, Taylor's law, and patchiness can be derived from common underlying mechanisms, and which ecological parameters determine the characteristic scales separating different spatial regimes.

\begin{figure}
    \centering
    \includegraphics[width=1\linewidth]{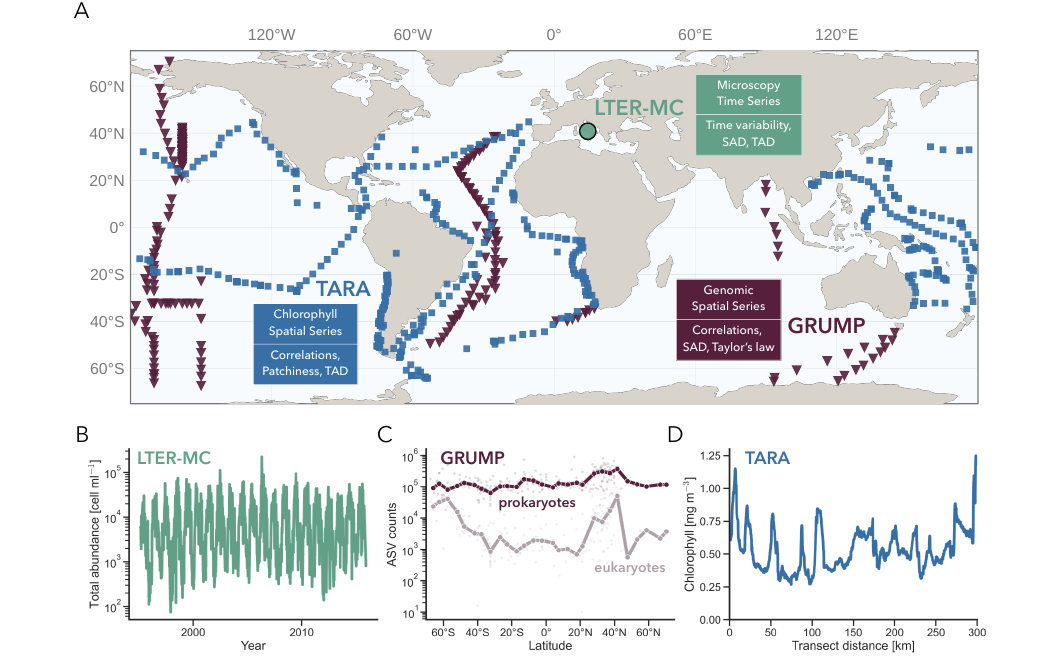}
    \caption{
    \textbf{Overview of analyzed datasets.} 
    (\textbf{A}): Geographic distribution of the datasets analyzed in this study. Green circles indicate the Long-Term Ecological Research MareChiara (LTER-MC) station, red triangles denote samples from the Global rRNA Universal Metabarcoding Plankton (GRUMP) database, and blue squares indicate transects from the Tara Pacific expedition and Tara Microbiome expedition. Boxes summarize the main data types and analyses considered. 
    (\textbf{B}): Example time series of total plankton abundance from the LTER-MC station (1995--2015).
    (\textbf{C}): Latitudinal distribution of total metabarcoding Amplicon Sequencing Variants (ASV) richness in the GRUMP dataset for prokaryotes and eukaryotes.
   (\textbf{D}): Example of spatial variation in chlorophyll concentration along a Tara transect. 
    See~\cite{methods} and the supplementary text.
    }
    \label{fig0}
\end{figure}

Here we develop an analytically tractable framework that links stochastic population dynamics at local scales to emergent spatial patterns of plankton diversity. The theory shows how fluctuations in per-capita growth, coupled by spatial diffusion, generate a characteristic correlation length and drive population variability across the seascape. From the same dynamics, we derive predictions for spatial correlations, species- and total-abundance distributions, spatial Taylor's law and plankton patchiness. We test these predictions across independent plankton datasets (see Fig.~\ref{fig0}) spanning long-term microscopy observations, global metabarcoding surveys and high-resolution chlorophyll transects, finding that distinct macroecological patterns can be traced back to the same underlying stochastic mechanism. By providing a common origin for patterns that have largely been studied separately, our results establish a quantitative bridge between local ecological fluctuations and large-scale marine biodiversity, offering a minimal baseline for identifying when additional biological or oceanographic processes are required.

\subsubsection*{Results}\label{sec:Results}
We first introduce a model for the spatial dynamics of plankton populations. Ideally, community dynamics would be described by tracking the abundance of every species across space and time. In practice, however, even long-term monitoring programs are limited to a subset of taxa (e.g.,~\cite{dAlcala2004_MareChiara}), making a comprehensive description of entire communities infeasible. 
An alternative strategy is to reduce the community-level description to the dynamics of a single focal species, assumed to be statistically representative of a typical community member. Under this approximation, the community is described by a neutral model with density-dependent effects. Without loss of generality, we consider the dynamics of the focal species in a one-dimensional system, represented as a chain of adjacent patches separated by a distance $u$.

\begin{figure}
 \centering
    \includegraphics[width=\linewidth]{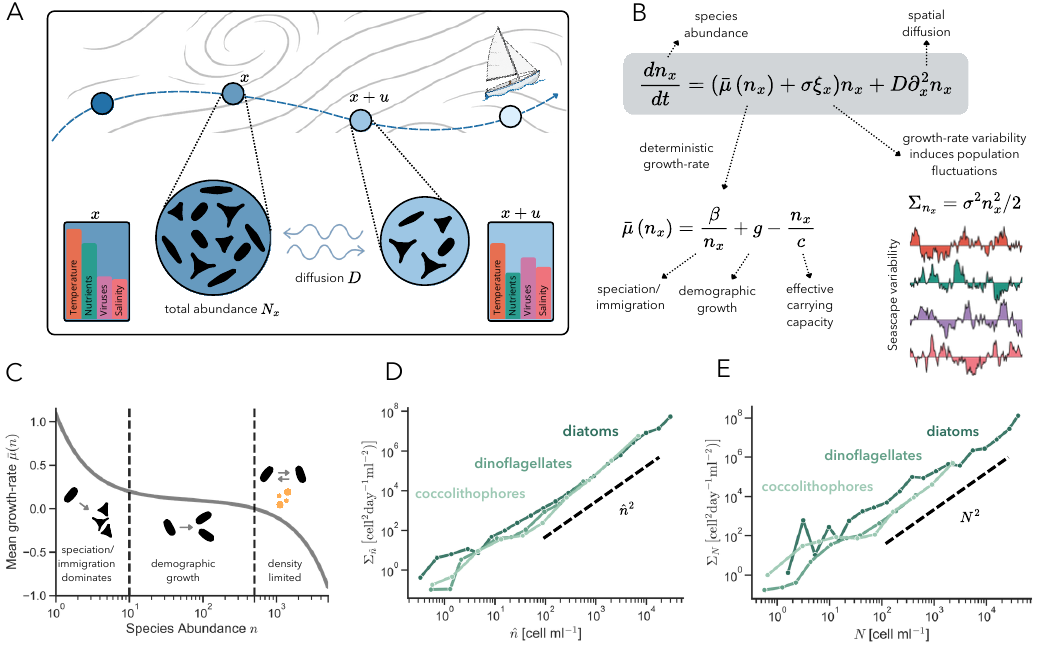}
    \caption{
    \textbf{Model summary.} 
    (\textbf{A}): Schematic representation of the seascape and a possible sampling transect (blue dashed line), with individual sampling locations shown as blue dots. Spatial and temporal heterogeneity arises from environmental variables such as temperature, nutrient availability, salinity, and viral abundance.
    (\textbf{B}): Schematic representation of the model equation and its ecological interpretation. The deterministic per-capita growth rate is decomposed into immigration/speciation, demographic growth, and density-dependent regulation, while stochastic growth-rate fluctuations induced by the seascape generate abundance fluctuations. The last term accounts for spatial dispersal.
    (\textbf{C}): Deterministic per-capita growth rate $\bar{\mu}$ as a function of abundance. Immigration dominates at low abundance, demographic growth at intermediate abundance, and density-dependent regulation at high abundance.
    (\textbf{D})~and~(\textbf{E}): Scaling of abundance fluctuations, $\Sigma_n$, from the LTER-MC dataset. Panel~D refers to focal species while panel~E to total abundance. Green shades denote different phytoplankton groups, and dashed lines show the theoretical prediction of the model. See~\cite{methods} for more details.
    }
    \label{fig:1}

\end{figure}

Within this framework, $n_x(t)$ indicates the density of focal species at time $t$ and location $x$, and the model parameters are understood as effective, namely, quantities that encode the aggregate outcome of the demography, the species interactions and the surrounding environment. This approach does not assume that the species are identical, but rather that their statistical properties are similar. The population density $n_x(t)$ is governed by the following equation
\begin{equation}\label{eq:model_intro}
    \frac{d n_x}{dt}=n_x \mu(n_x) +D \partial_x^2 n_x \qquad\text{with}\qquad \mu(n_x)=\bar{\mu}(n_x)+ \sigma\xi_x,
\end{equation}
where the per-capita growth rate is decomposed into a deterministic component,
$\bar{\mu}(n_x)=\beta\, n_x^{-1}+g-c^{-1}n_x$,
and a fluctuating component, $\sigma\xi_x$. Here, $\xi_x$ is a standard Gaussian white noise, delta-correlated in space and time~\cite{gardiner2009_stochastic}, and $\sigma$ controls the intensity of environmental fluctuations. By an appropriate rescaling of the system parameters~\cite{note1}, $n_x$ may be interpreted as the abundance measured along a transect of unit length ($u=1$). A sketch of the model is reported in panels A~and~B of Fig.~\ref{fig:1}.

The deterministic part $\bar{\mu}(n_x)$ captures three ecologically distinct mechanisms that operate across different abundance regimes. The first term, $\beta/n_x$, encodes the combined effects of immigration from a regional species pool or rare bloom events. This term dominates at low abundances, where populations are primarily maintained by recruitment rather than by local reproduction. The second, constant contribution $g$ describes intrinsic per-capita reproduction: each individual contributes offspring at a fixed rate, independently of local density, reflecting the replication rate characteristic of intermediate abundance regimes, where neither resource limitation nor top-down control is yet binding. Finally, the term $-n_x/c$ introduces a negative density dependence, parameterized by the carrying capacity $c$: as abundance increases, intra-specific competition for limiting resources and intensified grazing pressure by zooplankton collectively reduce the per-capita growth rate, providing the restoring force that prevents unbounded population growth. Together, these three terms generate the sigmoidal per-capita growth profile illustrated in panel~C of Fig.~\ref{fig:1}, transitioning from net positive growth at low-to-intermediate abundances to net negative growth at high abundances.

From an ecological standpoint, $\sigma\xi_x$ represents fluctuations in individual growth rates induced by the seascape~\cite{ottino2020_population, swartz2022_seascape}. Local environmental variability together with short-timescale biological processes are, therefore, modeled as stochastic fluctuations in the growth rate. This formulation is consistent with current marine ecosystem models~\cite{ottino2020_population,swartz2022_seascape,mallmin2024_chaotic,Mallmin2026_neutral_time,amer2026_spatiotemporal} and with the interpretation proposed by Campbell~\cite{Campbell1995_lognormal}.

The last term in Eq.~\ref{eq:model_intro} accounts for spatial dispersal induced by currents and mixing. Although oceanic transport is intrinsically complex and three-dimensional, several large-scale observables of plankton ecosystems are effectively controlled by transport along dominant flow directions. 
We therefore restrict the analysis to a one-dimensional geometry, which retains the essential transport mechanisms while allowing for analytical tractability. Furthermore, following the arguments of~\cite{Pasquero2005_differential}, the effective diffusion coefficient $D$ captures the combined action of turbulent mixing and unresolved advective processes at coarse-grained scales.

The stationary behavior of the model provides the distribution of local density, $P(n_x)$, which is the probability of finding a species with density $n_x$. In the absence of spatial diffusion, i.e., $D=0$, we can derive the following expression from Eq.~\ref{eq:model_intro}:
\begin{equation}\label{eq:local_dist}
P(n_x)
\propto
n_x^{-2+\frac{2g}{\sigma^2}}
\exp\left(
-\frac{2\beta}{\sigma^2 n_x}
-
\frac{2n_x}{c\sigma^2}
\right).
\end{equation}
This distribution, known as the Generalized Inverse Gaussian (GIG) distribution~\cite{Jorgensen1982_gig}, exhibits three distinct regimes that directly reflect the behavior of the deterministic growth rate $\bar{\mu}(n_x)$. At low abundances, immigration prevents extinction and sets a characteristic population scale $n_{\textrm{im}}=\beta/\sigma^2$, below which recruitment from outside the local community becomes dominant. At high abundances, top-down control defines a second characteristic scale, $n_{\textrm{top}}=c\sigma^2$, above which competition for resources and other density-dependent processes suppress further population growth. 

When $n_{\textrm{im}}\ll n_{\textrm{top}}$, the community experiences a broad intermediate-abundance regime in which populations can fluctuate on several scales without a preferred abundance. Reproduction dynamics dominate, yielding a power-law (first factor in Eq.~\ref{eq:local_dist}). When $n_{\textrm{im}}\simeq n_{\textrm{top}}$, stronger top-down control reduces species heterogeneity (e.g., due to resource limitation or low environmental variability). The SAD then becomes unimodal, closer to log-normal, with sizes clustering around $\sqrt{\beta c}$.

So far, we have focused on the dynamics of focal species and on the mechanisms shaping its relative abundance within the community. The same modeling framework can also be applied to the dynamics of total abundance, defined as the sum of the abundances of all species. This does not imply that total and relative abundances are governed by the same ecological processes. Rather, they can be described by the same mathematical structure, provided that the model parameters are interpreted according to the underlying ecological mechanisms. Under this interpretation, the deterministic dynamics of total abundance are described by logistic-type growth, in which a linear production term is limited by density-dependent saturation. External inputs, such as cross-flow transport or resource inflows, are naturally incorporated through the immigration term. The deterministic component of Eq.~\ref{eq:model_intro} therefore retains the same mathematical form, while its parameters acquire a different ecological interpretation. 
Likewise, the stochastic component of the per-capita growth rate represents fluctuations in environmental factors, including variability in production, transport, and resource availability. From now on, we will use $\hat{n}$ to denote the population of focal species, $N$ to denote the total abundance, and $n$ to represent all patterns that apply to both observables. 

A key ingredient of our modeling framework is the variability of the population growth rate. To characterize its dependence on abundance, we consider the quantity~\cite{Aguilar2025_limits,vankampen2004_stoch} 
\begin{equation}\label{eq:Sigma}
    \Sigma(n)=\left\langle \frac{\delta n_T^2(n)}{T}\right\rangle
\end{equation}
where $\delta n_T(n)= n_{t+T}-n$ is the change in abundance over a lag $T$, conditioned on the population being at $n$ at time $t$, and $\langle \cdot \rangle$ denotes the expectation value. The function $\Sigma$ quantifies the magnitude of abundance fluctuations induced by growth-rate variability. Under the stochastic model in Eq.~\ref{eq:model_intro}, from empirical data one expects to find a power-law scaling $\Sigma(n)\sim n^2$, which is the distinctive signature of local environmental fluctuations in  the growth rate. 

We verify the scaling for both the focal species abundance, $\hat{n}$, and total community population abundance, $N$, by analyzing the Long-Term Ecological Research MareChiara (LTER-MC) plankton time series.  Since we are dealing with local datasets, we omit the subscript $x$ for simplicity. We consider three groups of plankton: diatoms, dinoflagellates and coccolithophores. The results of our analysis are illustrated in Fig.~\ref{fig:1}. In Panels~D~and~E each point represents the geometric mean of $\delta n_T^2(n)/T$ for the corresponding initial population $n_t=n$, where $T=7~$days is the average sampling time. The dashed lines are the expected behavior of $\Sigma(n)\sim n^2$ (Eq.~\ref{eq:Sigma}), and we find excellent agreement between the data and the theory. We confirm empirically that population fluctuations are well described by the scaling $n^2$ for both focal species and total abundance. Details of the data analysis are provided in~\cite{methods} and the supplementary text (see Fig.~\ref{fig:Methods_S}).

A central finding of this analysis is that the exponent governing the noise scaling, $\Sigma(n) \sim n^2$, is conserved not only across the three plankton groups but also across levels of biological aggregation, holding equally for focal-species abundance and for total community abundance (Fig.~\ref{fig:1}~D-E). This indicates that the multiplicative structure of environmental noise in Eq.~\ref{eq:model_intro} is a general feature of the growth-rate fluctuations, which is independent of taxonomic identity or of whether the dynamics are considered at the scale of a single-species or at community level.

We highlight that our analysis neglects the effects of seasonality on the population growth rates. Indeed, we show in the supplementary text that the temporal dynamics of plankton abundances are characterized by relatively higher values during the summer months (Fig.~\ref{fig:Methods_seasonality_N}). The stochastic fluctuations which we have described above cannot capture this systematic seasonal pattern; instead, the seasonal variation should be explicitly incorporated in a temporal dependence of mean growth rate $\bar{\mu}$. However, as demonstrated in the supplementary text, when the time series is divided into monthly intervals, the assumption of a homogeneous $\bar{\mu}$ holds, indicating that our framework is applicable at appropriate temporal scales.

\subsubsection*{Spatial diffusivity and stochastic growth generate patchy population structures}

We now consider how spatial diffusion modulates stochastic population dynamics. In the model which we have introduced, both diffusion and local deterministic dynamics tend to homogenize the population: diffusion smooths spatial gradients, while density-dependent growth drives each site toward a common stable density. On the contrary, the fluctuations of the local growth rate counteract this tendency, generating spatial heterogeneity across neighboring regions. The interplay between these two mechanisms produces nontrivial spatial correlations and the emergence of patchy population structures.

We first study this effect by using the two-point correlation function: $C(x,y)=\langle n_xn_y\rangle - \langle n_x\rangle \langle n_y\rangle$. Assuming that the average population does not depend on the location of the patch and that $C(x,y)$ depends only on the distance $r=|x-y|$ between two patches, we find  (see~\cite{methods} and the supplementary text) the following expression:
\begin{equation}\label{eq:Correlation}
    C(r)=\frac{\tau \langle n^2\rangle \sigma^2}{4 \Gamma} \exp{\left(-\frac{r}{\Gamma}\right)},
\end{equation}
where $\tau$ is a characteristic time scale of the system and $\Gamma$ is the correlation length. The exponential profile of $C(r)$ differs from the linear decay generally observed for plankton community similarity~\cite{milici2016_bacterioplankton, Villarino2018_MALASPINA_corlength}. Nevertheless, as we shall show, the exponential trend perfectly describes the spatial correlations in terms of abundances. Moreover, from Eq.~\ref{eq:Correlation} we can estimate the population  variance by taking the prefactor, namely, $Var(n)=C(0)$.

Having characterized the two-point correlation structure of the model, we now move beyond pairwise statistics to ask how spatial diffusion changes the full distribution of local abundances. In the absence of diffusion, Eq.~\ref{eq:model_intro} reduces to a single-site process whose stationary distribution is a GIG, as established in Eq.~\ref{eq:local_dist}. For the full spatial stochastic dynamics, however, $P(n_x)$ cannot generally be obtained analytically from Eq.~\ref{eq:model_intro} as all patch sites are coupled.

To make analytical progress, we map the spatial stochastic differential equation onto an effective equation in which the effect of space is absorbed into a rescaling of the model parameters~\cite{Peruzzo2020_neutral_space} (for more details, see~\cite{methods} and the supplementary text). This approach works well for moderate values of the constant $D$. Applying this reduction to Eq.~\ref{eq:model_intro} yields the effective dynamics:
\begin{equation}\label{eq:results10}
\frac{dn}{dt}= \beta +g n-\frac{n^2}{c}+\sigma^\star n\, \xi, \qquad \sigma^\star = \frac{\sigma}{\sqrt{2\Gamma}}\quad,
\end{equation}
where now the parameters refer to the single patch population.  The effect of spatial diffusion enters implicitly through the rescaled noise amplitude $\sigma^\star$.  By construction, Eq.~\ref{eq:results10} reproduces the stationary mean and variance of the full spatial model, allowing us to analytically study the impact of space on the population distribution, rather than using only the numerical integration of the spatial equation.  For relatively larger values of diffusivity $D$, this approach is no longer appropriate and we leverage a different approximation.

There are several mathematical advantages and theoretical insights that we gain from Eq.~\ref{eq:results10}. First, under specific parameter limits (detailed in the supplementary text), we obtain a simple approximation for $\Gamma$:
\begin{equation}\label{eq:Gamma_approx}
    \Gamma\approx \frac{4D}{\sigma^2}\quad.
\end{equation}
The correlation length thus emerges from the balance between stochastic growth and spatial transport. Increasing diffusion enlarges the characteristic patch size and progressively homogenizes the system, whereas stronger growth-rate fluctuations enhance spatial heterogeneity.

Secondly, we expect to find two distinct regimes in the distribution of the populations (see~\cite{methods} and the supplementary text). In the first one, the local variance remains large enough that diffusion acts only as a perturbation, thus rescaling the noise component while leaving the deterministic part of the dynamics nearly unchanged. In this case, Eq.~\ref{eq:results10} provides a good approximation and dictates that the parameter $\sigma$ has to be simply rescaled, i.e. $\sigma^\star =\sigma/\sqrt{2\Gamma}$. Accordingly, we identify the power-law component of the population distribution as $P(n)\sim n^{-\lambda}$, where  $\lambda=2(1-g/{\sigma^{\star}}^2)$. In the second regime, diffusion is sufficiently large to suppress local population variability and Eq.~\ref{eq:results10} is no longer a good approximation of the full model. In this case, we find analytically (see~\cite{methods} and the supplementary text)  that populations are clustered around a characteristic scale and $P(n)$ is well approximated by a lognormal-like distribution. These theoretical regimes provide a mechanistic basis for recent empirical findings by Danling \emph{et al.}~\cite{danling2026_inv}, who showed that planktonic SADs can be grouped into lognormal-like and power-law-like distributions.

The transition between these two alternative approximations can be investigated analytically by using the power-law slope $\lambda$. More precisely, we find that for $\lambda\lesssim 1.5$, abundance fluctuations are relatively small and the distribution is close to lognormal. In contrast, for $\lambda \gtrsim 1.5$, the population dynamics show higher variations, diffusivity becomes less important, and the stationary distribution remains a GIG, yet with rescaled parameters. This transition is confirmed by numerical integration of Eq.~\ref{eq:model_intro} as shown in Fig.~\ref{fig:Methods}.

Our theoretical predictions are clearly recovered in the empirical data we analyzed.  To determine whether a distribution follows a lognormal or a GIG distribution, we employ the Bayesian Information Criterion~\cite{schwarz1978_estimating} by calculating $\Delta_\textrm{BIC}$~\cite{methods}: when $\Delta_\textrm{BIC}$ is positive, the sample is better represented by a GIG distribution rather than a lognormal, and the opposite for $\Delta_\textrm{BIC}$ negative.

\begin{figure}
   \centering
    \includegraphics[width=\linewidth]{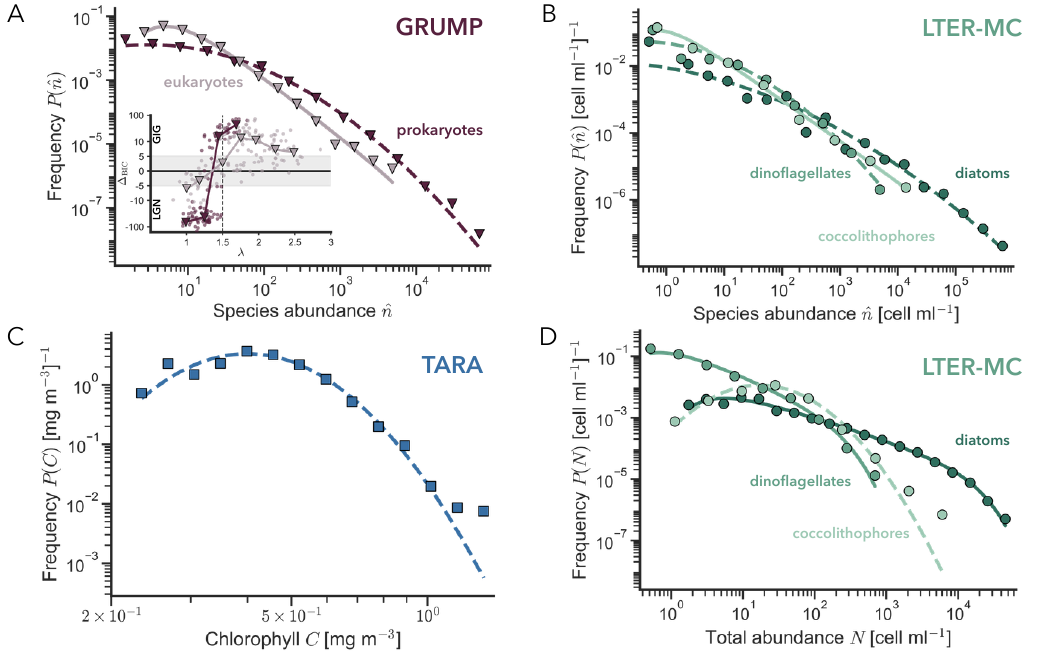}
    \caption{\textbf{Population distributions.} In all panels, solid curves denote the best-fit Generalized Inverse Gaussian (GIG) distribution, whereas dashed curves show the best-fit log-normal distribution. (\textbf{A}): Species abundance distributions (SADs) for representative eukaryotic and prokaryotic samples from the GRUMP dataset. The inset summarizes the transition between the GIG and log-normal regimes as a function of $\lambda$. Triangles indicate the median $\lambda$ values, and the gray region marks the interval where the two distributions are statistically indistinguishable according to the Bayesian Information Criterion ($|\Delta_{\mathrm{BIC}}|<5$). (\textbf{B}): Time-aggregated SADs for diatoms, dinoflagellates, and coccolithophores in the LTER-MC dataset. For coccolithophores, we have $\Delta_{\mathrm{BIC}}=0.08$, suggesting that lognormal and GIG are equivalent. (\textbf{C}~and~\textbf{D}): Examples of total abundance distributions. Panel~C shows the distribution of chlorophyll concentration from the Tara Microbiome/Tara Pacific dataset, whereas panel~D shows the total abundance distributions (TADs) for the LTER-MC dataset.
    }
    \label{fig:2}
\end{figure}

In Fig.~\ref{fig:2} we report the results of our analysis: panels~A~and~B show the SADs for two different datasets; panel~A refers to The Global rRNA Universal Metabarcoding Plankton (GRUMP) database. For the samples shown in panel~A, we recover a GIG distribution for eukaryotes ($\lambda=1.9$) and a lognormal for prokaryotes ($\lambda=1.1$). In the inset of panel~A, we show the predicted double behavior of $P(n)$: given the wide range of $\lambda$ values estimated in GRUMP SADs, we use these data to verify how the stationary distribution changes for different values of $\lambda$. For both prokaryotes and eukaryotes, we find that the distribution transition occurs around $\lambda=1.5$, confirming our expectations. In panel~B, we present the time-aggregated SADs for diatoms, dinoflagellates, and coccolithophores from the LTER-MC dataset, calculated from the total abundance of each species over the entire time series~\cite{methods}.  For every group, we find $\lambda<1.5$ and $\Delta_{\textrm{BIC}}\leq0$, confirming distributions with a shape that is more lognormal-like. Finally, in panels~C~and~D we present the Total Abundance Distributions (TADs), i.e., the distribution of the total population abundances. In detail, for panel~C we use chlorophyll concentration as a measurement of phytoplankton total abundance from the TARA dataset recently studied by Gray~\cite{Gray2025_chl_pathiness}.  In Fig.~\ref{fig:2} we consider a transect of $\sim1500$ samples distributed over $\sim300$ kilometers. We find that the chlorophyll distribution is lognormal for most of the transects analyzed (supplementary text). Thus, we recover for a spatial collection of samples the same statistical pattern found for a time series~\cite{Campbell1995_lognormal}, confirming that the spatial variability can be represented with a fluctuating growth rate. Instead, panel~D shows the TADs estimated from the species counts of the LTER-MC dataset. Here, we find $\lambda<1.5$ for each group. However, diatoms and dinoflagellates population distribution follow a GIG, thus deviating from the theoretical expectation. This divergence can be attributed to the seasonal effects. In the supplementary text,  we repeat our analysis using a monthly temporal resolution. For each month, we study the corresponding TAD, recovering the transition of the population distribution at $\lambda=1.5$ for each group (Fig.~\ref{fig:Methods_deltaBIC}). This confirms that once the heterogeneity caused by seasonality is removed, the model we propose successfully captures the population abundance variability.

\subsubsection*{From local to spatial variability}\label{sec:spatial}

\begin{figure}
    \centering
    \includegraphics[width=\linewidth]{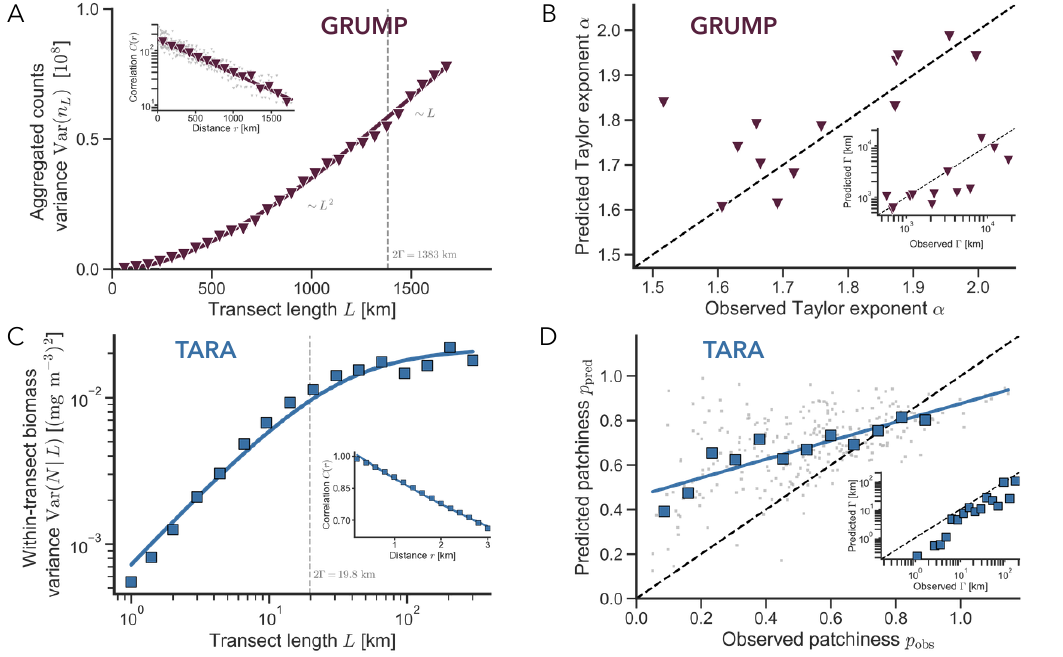}
    \caption{\textbf{Spatial patterns}.
    (\textbf{A}): Spatial Taylor's law for prokaryotic GRUMP samples from transect KM1906 near Hawaii. Triangles represent the data, and the solid line is the theoretical prediction from Eq.~\ref{eq:results13} using the fitted correlation length and local variance. The vertical dashed line marks the crossover scale $L=2\Gamma$, separating the correlated and de-correlated regions. The inset shows the spatial two-point correlation function, with triangular markers denoting the median correlation within distance bins and the red line the exponential fit.
    (\textbf{B}): Comparison across all transects between the observed Taylor's law exponent $\alpha$ and the theoretical prediction derived from the fitted correlation length. The inset compares the correlation lengths $\Gamma$ estimated from the empirical correlation function and from Eq.~\ref{eq:results13}.
    (\textbf{C}): Conditional variance $Var(n|L)$ as a function of $L$ for chlorophyll concentration in a representative Tara Microbiome--Tara Pacific transect. Dots represent the data, and the solid line is the fit of Eq.~\ref{eq:results14}. The inset shows the corresponding spatial correlation function.
    (\textbf{D}): Comparison between the observed patchiness exponent $p$ and the theoretical prediction from Eq.~\ref{eq:results14} over all samples. Gray dots denote individual samples, blue squares are median values within bins, and the blue line is the best linear fit. The inset compares the correlation lengths $\Gamma$ estimated from Eq.~\ref{eq:results14} and from the empirical spatial correlation function.}
    \label{fig:3}
\end{figure}

We now illustrate how diffusivity shapes spatial heterogeneity, focusing on two related quantities: the variance of the total population within a spatial window or transect of length $L$, namely $Var(n_L)$, where $n_L=\int_Ln_x dx$; and patchiness, a measure of how unevenly individuals are distributed across space.

The first captures how population fluctuations accumulate as the observation window is enlarged and typically scales as a power-law with respect to sample size $L$ (or, equivalently, in $\langle n_L\rangle$), following the empirical pattern known as Taylor's Law~\cite{taylor1961_aggregation, eisler2008_fluctuations}, $Var(n_L)\sim L^\alpha$. Integrating the correlation function given by Eq.~\ref{eq:Correlation} over the length of a transect~\cite{methods}, we find
\begin{equation}\label{eq:results13}
    Var\left(n_L\right)=\frac{\tau \langle n^2\rangle{\sigma}^2}{2}L\psi\left(\frac{L}{2\Gamma}\right)
\end{equation}
The function $\psi(z)$~\cite{methods} encodes how spatial correlations modulate the growth of variance with $L$, interpolating between two limiting behaviors: it grows linearly in $z$ when $z\ll 1$, while it saturates to 1 for $z\gg 1$. Therefore, $Var\left(n_L\right)$ shows a double regime set by the ratio between the transect length $L$ and the correlation length $\Gamma$: for $L\ll\Gamma$, neighboring abundances are strongly correlated and local fluctuations reinforce each other, so the variance grows superlinearly as $L^2$. For $L\gg\Gamma$, on the contrary, populations located in patches far more than $\Gamma$ fluctuate almost independently and the variance scales linearly with $L$, as expected from a standard central-limit argument.

The same correlation structure also governs patchiness, which is the spatial heterogeneity of the population. Following~\cite{mahadevan2002_biogeochemical}, the patchiness is quantified by the empirical variance of the local abundances evaluated across a transect of length $L$, which we denote $Var(n|L)$ (see~\cite{methods} and the supplementary text). Assuming a power-law relation  $Var(n|L)\sim L^p$, the exponent $p$ gives a compact summary of the dominant scale of spatial heterogeneity: small $p$ indicates that variability is concentrated at scales finer than the observation window $L$;  large $p$, instead, indicates that heterogeneity continues to grow as $L$ is enlarged, reflecting broad-scale spatial structure rather than fine-grained patchiness. Employing the scaling law defined by Eq.~\ref{eq:results13}, we obtain a functional relation between $Var(n|L)$ and the ratio $\Gamma/L$:  
\begin{equation}\label{eq:results14}
    Var(n|L)=Var(n)\cdot\left[1-\frac{2 \Gamma}{L}\psi\left(\frac{L}{2\Gamma}\right)\right]
\end{equation}
As with $Var(n_L)$, even $Var(n|L)$ has a double behavior governed by the ratio $L/\Gamma$: for $L\ll \Gamma$ there is a linear trend ($p=1$): in this case the observation window is too small to sample the full extent of spatially correlated patches. For $L\gg \Gamma$, $Var(n|L)$ saturates ($p=0$) as the transect becomes large enough to average over many independent patches, leaving no further heterogeneity to detect at larger scales.

These results show that spatial patterns of variance and patchiness interpolate between two limiting regimes governed by the ratio of correlation length $\Gamma$ over observation scale $L$. Our analysis establishes a direct link between these two patterns through their scaling exponents $\alpha$ and $p$ and the correlation length $\Gamma$, unifying two seemingly distinct empirical descriptors of spatial structure under a common mechanistic origin.

Our results are in agreement with empirical data. We analyze the spatial composition of the prokaryote community using the GRUMP dataset, which provides a higher resolution in terms of counts compared to the corresponding eukaryotic dataset~\cite{methods}. Panel~A of Fig.~\ref{fig:3} illustrates the spatial Taylor's law $Var(n_L)$ for the prokaryote samples. The variance of ASV abundance is plotted as a function of the spatial scale $L$ for an increasing number of samples combined according to spatial proximity. The line is the prediction from Eq.~\ref{eq:results13}, using $\Gamma$ estimated from the spatial correlation and fitting the local variance. In the inset, the two-point correlation function (supplementary text) is plotted as a function of geographical distance for the surface samples of prokaryotes of a representative transect. For most transects, the exponential decay is clearly visible, and the corresponding correlation lengths are of the order of 100 to 10,000 kilometers.  From the fit of $Var(n_L)$ we recover the two regimes predicted by the theory: for distances shorter than $2\Gamma$, samples are strongly correlated and the variance scales quadratically with $L$, while for larger distances samples behave as if they were independent, resulting in a linear scaling. Furthermore, using Eq.~\ref{eq:results13} we can relate $\Gamma$ to $\alpha$. For each transect, we estimate $Var(n_L)$ using the values of $\Gamma$ obtained from the empirical correlation. By interpolating this function with a power-law, we derive a theoretical prediction of $\alpha$. In panel~B, we report the comparison between $\alpha$ obtained by directly fitting a power-law relation and the theoretical values. We observe good agreement ($R^2=0.52$), which shows that this pattern is mainly driven by the behavior of the correlation function. Moreover, by fitting  Eq.~\ref{eq:results13}, we find an alternative estimate of $\Gamma$. The comparison with the $\Gamma$ values fitted by the correlation function is presented in the inset of panel~B, and the resulting covariance shows a remarkable agreement.

For the analysis of chlorophyll patchiness, we explored the TARA Pacific-Tara Microbiome dataset~\cite{Gray2025_chl_pathiness}. First, we divide the samples into consecutive transects of approximately $300~$km. For each of these, we take several subsets of length $L$ and calculate the variance of chlorophyll concentration within them, estimating $Var(n|L)$~\cite{methods}. In panel~C of Fig.~\ref{fig:2}, we report $Var(n|L)$ as a function of $L$ for a representative transect in the Pacific Ocean, together with the best fit of Eq.~\ref{eq:results14}. Two regimes are evident: samples separated by short distances remain strongly correlated and the variance scales linearly with $L$; at larger distances, the samples become uncorrelated and the variance saturates. In the inset we show the spatial correlation of the same transect. Similarly to GRUMP, we recover an exponential decay but with markedly shorter correlation lengths, namely $\Gamma\approx 6.9~\text{km}$.  Finally, in panel~D, we report the values of $p_{\textrm{pred}}$ derived employing Eq.~\ref{eq:results14} and $p_{\textrm{obs}}$ measured by fitting a power-law relation. We note that the theoretical patchiness overestimates the observed values. This can be attributed to the transects within the oligotrophic regions. As we demonstrate in~\cite{methods} and in the supplementary text, some transects are characterized by low chlorophyll concentrations and appear to be more patchy with $p_{\textrm{obs}}$ further de-correlated with $\Gamma$ (see Fig.~\ref{fig:SM_patch}-C). These deviations are compatible with additional spatially structured processes, including localized vertical fluxes~\cite{Mahadevan2016_impact} and topographic upwelling from the island mass effect~\cite{gove2016_near}. Such processes interfere with spatial dispersal decoupling patchiness and spatial correlations. Our analysis indicates that such behaviors cannot be quantified by our model, suggesting the need to develop an alternative description for oligotrophic zones (see Fig.~\ref{fig:SM_patch} and supplementary text for more details).

We emphasize that the spatial patterns described here depend solely on the correlation function, from which we derive Eqs.~\ref{eq:results13}~and~\ref{eq:results14}. The theoretical model establishes a direct connection between the spatial scaling exponents ($\alpha$ and $p$) and the underlying biological processes.

Indeed, from Eq.~\ref{eq:results14} we can prove (supplementary text) that the patchiness exponent scales as $p\propto\log \Gamma\propto \log \tau$, the latter relation following directly from the definition of the correlation length. Patchiness therefore shows a direct connection with the characteristic temporal scale of the system, which, in terms of total abundances, represents the time necessary for the population to attain the equilibrium size after a perturbation.  This scaling law coincides with the one identified by Mahadevan \emph{et al.}~\cite{mahadevan2002_biogeochemical}, supporting the role of environmental stochasticity in generating spatial population heterogeneity.

\subsubsection*{Discussion}

Our results show that several prominent patterns of plankton variability that are usually studied separately can emerge from some underlying spatial stochastic mechanisms. By coupling multiplicative fluctuations in per-capita growth rate and effective spatial dispersal, we relate local abundance fluctuations to species- and total-abundance distributions to spatial heterogeneity patterns, namely Taylor's law and patchiness. Central to this connection is an emergent correlation length, $\Gamma$, which links the strength of local environmental variability to the spatial scale over which populations remain dynamically coupled. Rather than reproducing each macroecological pattern through a separate mechanism, the model therefore provides a unified conceptual framework of the seascape in which distinct observables arise as different manifestations of the same underlying dynamics.

Considerable interdisciplinary effort has been devoted to developing process-rich models of marine plankton ecosystems~\cite{follows2007_emergent,Sharoni2026_remodelling}. By incorporating detailed biological, physical, and biogeochemical mechanisms, such models can reproduce a broad range of observations and provide valuable system-specific predictions~\cite{tagliabue2023_oceans,flynn2025_more}. Their complexity, however, can make it difficult to determine which processes are necessary for a given macroscopic pattern and which instead reflect the specific ecological context being modeled. Because different mechanistic descriptions may reproduce the same large-scale observations~\cite{Purves2005_ecological}, agreement with data alone does not necessarily identify the minimal processes responsible for their emergence. Minimal models provide a complementary perspective by asking whether apparently complex ecological patterns can be explained from a small set of shared dynamical ingredients. Following this approach, our framework combines density-dependent regulation, effective spatial dispersal, and stochastic fluctuations in per-capita growth rate (Eq.~\ref{eq:model_intro}) while remaining sufficiently tractable to derive the resulting spatial statistics analytically.

A first ingredient of our study is the multiplicative structure of population fluctuations, previously suggested in temporal and theoretical descriptions of ecological dynamics~\cite{Campbell1995_lognormal,swartz2022_seascape,Mallmin2026_neutral_time,amer2026_spatiotemporal}. This assumption is directly testable since it implies that the abundance variance scales quadratically with abundance (Eq.~\ref{eq:Sigma}). The LTER-MC time series supports this scaling across diatoms, dinoflagellates, and coccolithophores and, notably, at two levels of biological aggregation: both focal-species and total-community abundance exhibit the same scaling, though with different effective parameters. This suggests that multiplicative growth-rate variability may represent a coarse-grained property of plankton population dynamics that persists across taxonomic and organizational scales. The evidence is nevertheless obtained from a single coastal site sampled at approximately weekly intervals and therefore does not establish that the same scaling holds universally in open-ocean communities. To our knowledge, however, this provides the first empirical demonstration of this fluctuation scaling in marine plankton.

When these local fluctuations are coupled through dispersal, their balance generates the characteristic spatial scale predicted by the theory. Within the approximation developed here, this is quantified by the parameter $\Gamma\approx 4D/\sigma^2 $ (Eq.~\ref{eq:Gamma_approx}), so that stronger effective diffusion extends spatial correlations whereas stronger variability in population growth shortens them. The exponential decay predicted for abundance correlations (Eq.~\ref{eq:Correlation}) is consistent with the GRUMP transects, but the importance of $\Gamma$ extends well beyond pairwise correlation. The same quantity controls how spatial coupling regulates local fluctuations, influences the form of abundance distributions, and determines the crossover scales of Taylor's law and patchiness. The correlation length therefore acts as a key metric that translates the balance between stochastic growth and dispersal into multiple macroscopic properties of the seascape.

This connection is particularly evident in the stationary abundance distributions. In the weakly homogenized regime, local fluctuations remain sufficiently strong for the population distribution to retain a broad GIG form, including an extended power-law-like range at intermediate abundances. As the homogenizing effect of dispersal becomes stronger relative to local growth-rate variability, fluctuations are suppressed and the distribution crosses over toward a lognormal-like regime. The theory identifies this transition analytically through the exponent $\lambda$ , rather than introducing the two distributional forms as independent phenomenological alternatives. The empirical data are broadly consistent with this prediction: GRUMP communities span both regimes, whereas chlorophyll concentrations along most Tara transects fall in the lognormal-like regime. In LTER-MC, pooled total-abundance distributions show departures from the theoretical association, but the expected relation is largely recovered after stratification by month. These results suggest that GIG and lognormal-like abundance distributions may represent different dynamical regimes of a common stochastic process rather than unrelated empirical functions~\cite{williamson2005_lognormal,mcgill2006_empirical}.

The same unifying framework extends to spatial fluctuation properties. Both spatial Taylor's law and patchiness follow analytically from the two-point correlation function and are therefore controlled by the ratio between the observation scale $L$ and the intrinsic correlation length $\Gamma$. For Taylor's law, the variance of integrated abundance grows quadratically with $L$ for scales shorter that $\Gamma$ because neighboring populations fluctuate coherently, and crosses over to linear growth for larger $L$ when distant regions become effectively independent (Eq.~\ref{eq:results13}). Patchiness exhibits a similar behavior: the variance among local observations initially increases with $L$ and then saturates once the sampling window exceeds the correlation scale (Eq.~\ref{eq:results14}). Taylor's law, patchiness, and spatial correlations are therefore not independent properties; they encode different aspects of the same underlying correlation structure. The close correspondence with GRUMP data, and the broader agreement with Tara chlorophyll transects, support this interpretation across substantially different spatial scales and observational modalities.

The departures from the theory are also informative because they identify where additional ecological or physical structure becomes relevant. In Tara, the association between patchiness and the predicted correlation structure weakens in oligotrophic transects, where local and heterogeneous events (e.g., vertical nutrient supply or localized transport) cause the homogeneity assumption underlying our model to fail. In LTER-MC, the improved correspondence obtained after stratifying observations by month is consistent with seasonal non-stationarity in the effective growth parameters. Neither result identifies a unique missing mechanism, but both indicate conditions under which the assumptions of spatially homogeneous and temporally stationary effective parameters become insufficient. The one-dimensional geometry and effective-diffusion description likewise do not explicitly represent anisotropic advection, transport barriers, or vertical exchange~\cite{Mahadevan2016_impact}, while the neutral approximation captures shared statistical behavior without resolving taxon-specific interactions or responses. These limitations define the domain of the present theory: systematic deviations from its predictions can be used to identify ecological regimes in which additional mechanisms are required.

This perspective suggests a broader role for the framework as a quantitative baseline for marine macroecology. Because the characteristic spatial scale is analytically related to effective diffusivity and growth-rate variability, independent estimates of $D$ and $\Gamma$ could, subject to the model assumptions, constrain the effective fluctuation amplitude  and permit comparisons across regions, seasons, or environmental regimes. The same framework may offer a possible route towards future climate applications. Changes in ocean circulation may influence the effective diffusivity $D$~\cite{Abernathey2013_global, zhang2023_global}; shifts in stratification and mixing~\cite{li2020_increasing, cheng2025_ocean} could be captured by the variability represented by $\sigma$; and evolving thermal and nutrient regimes~\cite{Kwiatkowski2020_twenty, moore2018_sustained} may modify the mean population growth. Because these changes are expected to have distinct, sometimes opposite, regional signs, climate-driven change in plankton spatial structure is likely to be markedly regional rather than a single global trend.

More generally, our results show that local stochasticity and spatial transport can jointly generate multiple patterns of plankton macroecology driven by  a common emergent spatial scale. This provides a mechanistic link between local population fluctuations and large-scale statistical regularities across marine seascapes. Although developed for plankton, the same framework may apply to other ecological systems in which dispersal interacts with multiplicative population fluctuations~\cite{Kalyuzhny2015_neutral_theory,Grilli2020_macro_model}. Testing whether this connection persists across ecosystems, and identifying structured deviations from it, provides a natural direction for extending the framework beyond marine communities.


\clearpage 

%
\bibliography{biblio} 
\bibliographystyle{sciencemag}


\section*{Acknowledgments}
The authors thank Amos Maritan, Davide Bernardi, and Javier Aguilar for insightful discussions.\\
The site LTER-MC (DEIMS.ID: https://deims.org/0b87459a-da3c-45af-a3e1-cb1508519411) belongs to the Long Term Ecological Research national and international networks: LTER-Italy\\ (https://ror.org/05ma8mw15), LTER-Europe and ILTER. We thank the LTER-MC team for their support. \\
S.~S., S.~A., D.~I., G.V.~V. and K.~S. received funding from the European Union’s Horizon Europe research and innovation program under grant agreement No. 101059915 (project BIOcean5D) and No. 862923 (project AtlantECO). Views and opinions expressed are however those of the authors only and do not necessarily reflect those of the European Union. Neither the European Union nor the granting authority can be held responsible for them.

\section*{Supplementary Materials}


\subsection*{Methods}

\subsubsection*{Logistic model with stochastic fluctuations}\label{sec:logistic_model_units}

The deterministic growth rate $\bar{\mu}(n_x)$ in Eq.~\ref{eq:model_intro} can be seen as an approximation of a general per-capita growth rate. Indeed, we can expand such a rate as a series of $n_x$, $\bar{\mu}(n_x)=\sum_k a_k n_x^{-k}$, with $a_k$ some constants, and $k$ can be both positive and negative. This approach follows the interpretation proposed by Volkov \emph{et al}.~\cite{volkov2005_density}, who expanded the per-capita birth and death rates in terms of $1/n$ to the first order. In this work, we extend this expansion including coefficients $k=-1,0,1$. Therefore, $a_{-1}=-1/c$, $a_{0}=g$ and $a_1=\beta$. 

The stochastic component $\sigma\xi$ represents the fluctuations given by both interactions with environment and biological processes.  This approach has recently been suggested in~\cite{ottino2020_population, swartz2022_seascape, mallmin2024_chaotic, Mallmin2026_neutral_time, amer2026_spatiotemporal} to capture the properties of the structure of the planktonic community.  However, the one-dimensional case involving this growth-rate noise term coupled with diffusivity was not explicitly addressed, nor was there any systematic analysis of the spatial patterns arising from the coupling between the stochastic forcing and the diffusion coefficient $D$. 

We emphasize that Eq.~\ref{eq:model_intro} refers to the dynamics of a population density; therefore, to obtain an abundance, we must integrate $n_x$ over a length. Considering patches of size $u$, the abundance associated with a single sample (see Fig.~\ref{fig:2}) can be written as $n=\int_{0}^un_xdx$.  For $\Gamma\gg u$, we have $n_x\approx n/u$, then $Var(n)\approx u^2Var(n_x)$. 

The system parameters can be properly rescaled in units of patch size $u$. The constant contribution to the growth rate $g$ and the diffusivity coefficient $D$ remain unchanged after scaling, whereas the other parameters are rescaled as following:
\begin{equation}
    \beta\to\beta u,\,\qquad c\to c u,\,\qquad \sigma\to \sigma u^{-1/2}.
\end{equation}
 In this way, the parameters $\beta$, $g$, and $c^{-1}$ have units of $\text{time}^{-1}$, whereas $\sigma$ has units of $\text{time}^{-1/2}$. Eq.~\ref{eq:model_intro} thus becomes the stochastic equation for local abundance. 
Since the spatiotemporal patterns of density $n_x$ and local abundance $n$ are equivalent, we interchange the two notations and set $u=1$.

\subsubsection*{Spatial scaling laws}\label{subsec:spatialscaling}
The core of spatial pattern analysis is the derivation of the two-point correlation function $C(r)$ (Eq.~\ref{eq:Correlation}), which we outline here. 

From Eq.~\ref{eq:model_intro}, it is possible to derive (see supplementary text) an equation for the time derivative of $C(r)$, that is,
\begin{equation}\label{eq:results5}
    \partial_t C(r)=2 g C(r)-\frac{U(r)}{c}+ 2D\partial_r^2 C(r)+\sigma^2 \langle n^2\rangle \delta(r),
\end{equation}
where $\delta(r)$ is the Dirac delta function and we define $U(r)=\langle n_xn_{x+r}^2\rangle+\langle n_{x+r}n_x^2\rangle-2\langle n\rangle\langle n^2\rangle$. 
$U(r)$ incorporates the non-linearity of the system resulting from the quadratic death rate. Yet, this term can be linearized as a function of $C(r)$. Indeed, for  small fluctuations and within the stationary regime,  we can write $\langle n_xn_{x+r}^2\rangle\approx\langle n_xn_{x+r}\rangle\langle n\rangle$, then $U(r)\approx 4\langle n\rangle C(r)$.  
This approximation is valid to leading order; to obtain a more accurate linearization, we proceed as follows. First, we set $U(r)=A+B\,C(r)$ and observe that $U(0)=2\langle n^3\rangle-2\langle n\rangle \langle n^2\rangle$ and $\lim_{r\to\infty}U(r)=0$. Within our linear ansatz, these relations become $U(0)=A+B\,Var(n)$ and $\lim_{r\to \infty}U(r)=A$. Imposing the equality, we derive an explicit formula for $U(r)$, that is, $U(r)=2c(\tau^{-1}+g)C(r)$, where the characteristic time scale $\tau$ reads
\begin{equation}
    \tau=\left(\frac{\langle n^3\rangle - \langle n^2\rangle\langle n\rangle}{c Var(n)}-g\right)^{-1}
\end{equation}
By inserting this expression into Eq.~\ref{eq:results5}, at stationarity we obtain
\begin{equation}\label{eq:sub_spatial_2}
    -\frac{1}{\tau}C(r)+D\partial_r^2 C(r)+\frac{\sigma^2}{2} \langle n^2\rangle \delta(r)=0,
\end{equation}
whose solution is given by Eq.~\ref{eq:Correlation}  (see the supplementary text and~\cite{Peruzzo2020_neutral_space}).

\subsubsection*{Self consistent relations: from the local distribution to the correlation length}\label{subsec:correlations}
We derive Eq.~\ref{eq:results10} from the scaling law Eq.~\ref{eq:results13}. First, note that in the homogeneous, noise-free case, diffusivity does not affect the stationary behavior of the system. This suggests that, at least for sufficiently small $D$, the core per-capita growth rate $\bar\mu(n)$ can be assumed to remain unchanged upon introducing a stochastic component. We therefore look for a new noise term $\sigma^\star$ that incorporates the diffusive effects such that, for any $x$, the quasi-stationary dynamics is described by a logistic equation like Eq.~\ref{eq:results10}. Keeping $\sigma^\star$ as a variable, the population variance reads as
\begin{equation}\label{eq:sub_corr_1}
    Var(n)=\frac{\tau {{\sigma}^\star}^2}{2}\langle n^2\rangle.
\end{equation}
From Eq.~\ref{eq:model_intro}, we observe that the variance can be derived from the two point correlation function: $Var(n)=C(0)$ when $\Gamma\gg1$. Hence, using Eq.~\ref{eq:Correlation}, we have
\begin{equation}\label{eq:sub_corr_2}
    Var(n)=\frac{\tau \sigma^2}{4 \Gamma}\langle n^2\rangle
\end{equation}
By imposing the equality between Eqs.~\ref{eq:sub_corr_1}~and~\ref{eq:sub_corr_2}, we arrive at the expression for $\sigma^\star$ reported in Eq.~\ref{eq:results10}.

Finally, we use this effective noise to simplify to obtain the expression for $\Gamma$ given by Eq.~\ref{eq:Gamma_approx}.  Starting from the effective dynamics given by Eq.~\ref{eq:results10}, we can compute the moments of the population abundance explicitly. By construction, the first two moments derived from Eq.~\ref{eq:results10} match exactly those of the full model in Eq.~\ref{eq:model_intro}. Assuming that the effective model also provides a reasonable approximation of the third moment, we arrive at the expressions for $\tau$ and $\Gamma$, and then to Eq.~\ref{eq:Gamma_approx}. A detailed derivation is provided in the supplementary text.

\subsubsection*{Patchiness}\label{subsec:patchiness}
The empirical variance $Var(n|L)$ defined in Sec.~\ref{sec:spatial} can be written explicitly as 
\begin{equation}\label{eq:sub_pat_1}
     Var(n|L)=\left\langle \int_{L}dx\frac{ \left(n_x -\int_Ldy\, n_y/L\right)^2}{L} \right\rangle
\end{equation}
where $L^{-1}\int_L n_xdx$ is the empirical mean of the population density for a sample of size $L$. Therefore, for as $L$ becomes large, this mean converges to the expectation values $\langle n\rangle$. Given that the patch has unit size, Eq.~\ref{eq:sub_pat_1} also corresponds to the empirical variance of local abundance. 

Generally, spatial heterogeneity is studied by decomposing the empirical variance into a Fourier series, to express it as a function of wavenumbers (see, for example,~\cite{Mackas1979_spectral, abraham1998_generation}). Following this spectral analysis, patchiness is quantified by examining the log-log scaling relationship between the variance decomposition and the wavenumbers, in order to identify the relative contributions of the various spatial fluctuations. In~\cite{mahadevan2002_biogeochemical} and more recently in~\cite{Gray2025_chl_pathiness}, patchiness was proposed to be measured using the scaling $Var(n|L)\propto L^p$. As demonstrated in~\cite{Gray2025_chl_pathiness}, the exponent $p$ is strongly correlated with the standard power-spectrum slope; nonetheless, $p$ is a better measure of patchiness when the spatial series is uneven. Moreover, as demonstrated in this work, it enables analytical studies. 

From Eq.~\ref{eq:sub_pat_1} and following the steps discussed in the supplementary text, the empirical variance becomes $Var(n|L)=Var(n)-Var(n_L)L^{-2}$. Then, using Taylor’s law (Eq.~\ref{eq:results13}), we obtain Eq.~\ref{eq:results14}.

\subsubsection*{Double regime of the probability distribution}\label{sec:double_regime}
To show how diffusion affects local density fluctuations, consider the following variable: $Y_x=\log\left(n_x/\langle n\rangle \right)$. In Campbell’s work~\cite{Campbell1995_lognormal}, $Y_x$ is assumed to follow a Gaussian distribution. Indeed, considering only the intermediate abundance regime in our model (when abundances are neither very small nor very large), we have $n_x(t)\sim \exp{(g t+\sigma \xi_t t)}$. The argument of the exponential corresponds to $Y_x+\log(\langle n\rangle)$, which is a Gaussian variable. To explore what happens when using $\bar{\mu}(n)$ (Eq.~\ref{eq:model_intro}),  we employ the Ito’s calculus (see~\cite{gardiner2009_stochastic} and the supplementary text). Denoting by $\Gamma_Y$ the characteristic length scale of $Y_x$ and by $\tau_y$ its characteristic time scale, we introduce the following rescaled quantities: $X=x/\Gamma_Y$ and $T=t/\tau_Y$. The spatial correlations are modulated by $\epsilon=[\tau_{Y}/(2\Gamma_Y)]^{1/2}\sigma$, which is the local standard deviation of $Y_x$ damped by the dispersal terms. Following the supplementary text, we can show that for large correlations and $\lambda<1.5$, the following equation holds:
\begin{equation}\label{eq:methods_double}
    \partial_T Y_{X}= - Y_{X}+ \partial_{X}^2 Y_{X}+\epsilon \xi_X(T) +\mathcal{O}( \epsilon^2)
\end{equation}
As the solution of Eq.~\ref{eq:methods_double} is a Gaussian process at leading order, $n$ is lognormally distributed. For $\lambda>1.5$, the lognormal approximation becomes inaccurate (supplementary text) and higher-order contributions to Eq.~\ref{eq:methods_double} must be included. In this case, local abundance fluctuations substantially exceed those we have in the lognormal regime, and diffusion enters only as a perturbation. An analogous dynamical behavior was investigated in~\cite{Peruzzo2020_neutral_space} and, following their approach, we incorporate the effect of spatial dispersal by replacing $\sigma$ with $\sigma^\star$, which yields Eq.~\ref{eq:results10}. These approximations have been always validated by numerical simulations (Fig.~\ref{fig:Methods}).

\subsubsection*{BIC analysis}\label{Methods:BIC}
To test the double behavior of the abundance distributions as a function of slope $\lambda$, we use the Bayesian Information Criterion. For each fit of a statistical distribution function to an empirical set of samples, we define the following quantity
$\textrm{BIC} = k \ln(\omega) - 2 \ln(\mathcal{L})$, with $k$ being the number of parameters of the theoretical distribution function to be estimated, $\omega$ the sample size and $\mathcal{L}$ the likelihood associated with the fit. When comparing two different models,  the distribution with the lower $\textrm{BIC}$ is preferred. Therefore, defining $\Delta_{\textrm{BIC}}=\textrm{BIC}({\text{LNM}})-\textrm{BIC}(\text{GIG})$ the difference of the $\textrm{BIC}$ evaluated fitting a log-normal or a GIG respectively, the sample is better represented by a GIG rather than by a log-normal distribution when $\Delta_{\textrm{BIC}}>0$, and the opposite for $\Delta_{\textrm{BIC}} < 0$.

\subsubsection*{MareChiara: Dataset}
The LTER-MC sampling site (40$^{\circ}$48.5'N, 14$^{\circ}$15'E) is located in the Gulf of Naples (Tyrrhenian Sea, western Mediterranean Sea). Samples have been collected since January 1984, with one major interruption from August 1991 to February 1995. Sampling was conducted fortnightly until 1991 and weekly from 1995 to the present. Phytoplankton samples were collected at a depth of 0.5 m using Niskin bottles, immediately fixed, and stored at 4$^{\circ}$C. Cell counts were performed according to the Utermöhl method, using an inverted light microscope at 400× magnification, following sedimentation of variable volumes of seawater (1–100~mL). Abundance data were obtained for approximately 320 taxa by examining 1-2 transects of a sedimentation chamber. The actual volume of seawater examined ranged from 0.02 to 1.52~mL.  Taxonomic identifications have been regularly intercalibrated among three experienced operators over the study period, with some having been involved since the start of the LTER program. In this work, we consider only taxa that are identified up to the species level. 

\subsubsection*{MareChiara: SAD and TAD}

Since most species are absent during a large portion of the sampling period, we aggregated the abundance of each species across the entire time series to obtain sufficient non-zero values for the SAD analysis. Defining $n^\alpha(t)$ as the abundance of species $\alpha$ at time $t$, the time-aggregated abundance is:
\begin{equation}
     n^\alpha=\sum_t n^\alpha(t)
\end{equation}
where the sum runs over the full time series. Consequently, the SADs shown in Fig.~\ref{fig:2} represent time-aggregated species abundances, computed as normalized histograms of $n^\alpha$ across all species $\alpha$.
For the study of TADs, we consider all values of the total abundances over all species of individual groups and use these to study the total abundance distributions. Therefore, TADs are the normalized histograms of the total counts across the time series. 

The results shown in Fig.~\ref{fig:2} do not include seasonal effects. Indeed, as discussed in detail in the supplementary text, by performing the $\Delta_{\textrm{BIC}}$ analysis on the aggregated, we cannot recover the expected transition between lognormal and GIG distributions for $\lambda\approx1.5$. Nonetheless, as illustrated in Fig.~\ref{fig:Methods_Sigma_n}, by examining the abundance distributions while dividing the time series by month, we reduce the seasonal heterogeneity  and observe the predicted transition.

\subsubsection*{MareChiara: stochastic noise}\label{SM:MC_noise}
Here we show the analysis of the noise amplitude  $\Sigma(n)$ from the MareChiara time series. First, we study the focal species behavior. For each bin of $\hat{n}$ of the distribution of abundances over species and time we look for the days at which a single species abundance falls within this bin.  We then take the abundance at time $t+T$, where we choose $T\in [5,9]~$days to represent a time resolution of one week. Then we collect a set of values $\delta n_T^2(\hat{n})/T=(\hat{n}_{t+T}-\hat{n})^2 /T$ and find $\Sigma(\hat n)$ taking the geometric mean of these values. We consider the geometric mean as an estimator of the expectation value of Eq.~\ref{eq:Sigma} because the empirical data are located in logarithmic scale. 

For the amplitude noise of the total abundances, we select a bin of $N$ from the TAD, and look for the days $t$ at which the total abundance of the group under analysis falls within that bin. Similarly to the focal-species analysis, we obtain a set of $\delta n_T^2(N)/T = (N_{t+T}-N)^2 /T$, and estimate $\Sigma(N)$ as their geometric means. 

Further information on the noise analysis is given in the supplementary text, together with the discussion on seasonal effects. As shown in Figs.~\ref{fig:Methods_S},~\ref{fig:Methods_Sigma_n}~and~\ref{fig:Methods_Sigma_N}, the scaling expected from environmental noise is validated for both $\hat{n}$ and $N$, across all months and for each species group.

\subsubsection*{GRUMP: dataset}
The GRUMP database combines DNA samples collected during several expeditions between 2003 and 2020. Although sampling procedures and DNA extraction protocols varied among expeditions \cite{McNichol2025_GRUMP}, the metabarcoding sequencing protocol has been standardized to allow direct comparisons between samples from different expeditions. We downloaded the GRUMP data version 1.3.5 from the Zenodo repository at \url{https://zenodo.org/records/15446776}.  DNA from unfractionated ($> 0.2~\mu m$) seawater samples was amplified using the 515Y/926 R universal three-domain rRNA gene primers, simultaneously quantifying the relative abundance of amplicon sequencing variants (ASVs) from bacteria, archaea, eukaryotic nuclear 18S, and eukaryotic plastid 16S. This procedure makes it possible to directly compare relative abundances of ASVs between samples and biological domains. On average, bacteria contributed $71\%$, eukaryotes $19\%$, and archaea $8\%$ to rRNA gene abundance, which explains why the patterns are much cleaner for the bacteria 16S data.

First, we consider only surface samples (depth smaller than 20 meters). Next, we make sure that the transects belonging to the different expeditions are as regular and continuous as possible by avoiding gaps and maintaining a regular spacing between the samples. When necessary, different surface samples taken at the same sampling station are combined into one \emph{spatial} sample. We used the raw sequencing counts but tested the impact of quality corrections, which were negligible.

\subsubsection*{GRUMP: spatial patterns}

For each transect, we estimate the correlation length by interpolating the spatial correlation function with Eq.~\ref{eq:Correlation}. 
In particular, since species are considered independent in the neutral context, the expectation value $\langle n_xn_{x+r}\rangle$ (and then $C(r)$) is estimated by taking the average across all sampled species. To analyze the spatial Taylor’s law, for each transect we group the samples  into subgroups of length $L$, and compute the variance for each group to estimate $Var(n_L)$. A detailed discussion on the fit of $\Gamma$, the analysis of the Taylor's exponent and the $\Delta_{\textrm{BIC}}$ test are left in the supplementary text.

\subsubsection*{Tara Chlorophyll: dataset}

Gray et al. \cite{Gray2025_chl_pathiness} compiled an extensive dataset of chlorophyll samples from the Tara Pacific \cite{Gorsky2019_Tara_Pacific} and Tara Microbiome \cite{Pesant2015_Tara_sampling} expeditions. All in-situ measurements are from the SV Tara underway Continuous Surface Sampling System which pulls seawater from 2 m depth. The chlorophyll concentrations are calculated based on optical data derived from particulate absorption and attenuation measured with a Seabird Scientific ACs. A $0.2~\mu m$ filter cartridge was connected to the system, and the flow was automatically redirected to measure the properties of the filtered seawater for ten minutes every hour. The samples were taken at intervals of $200$~meters, approximately one minute apart.

\subsubsection*{Tara Chlorophyll: spatial variability}
Patchiness is analyzed using the method proposed in~\cite{Gray2025_chl_pathiness}. Having defined a maximum spatial scale ($300~$kilometers in our case), we examine how the empirical variance of chlorophyll concentration scales with the length of the transect under analysis, i.e. with the number of samples. For each transect, we take 16 smaller lengths $L$, distributed logarithmically between $1~$kilometers and $300~$kilometers. After identifying all uninterrupted transects of $300~$kilometers, we consider sub-transects of various lengths $L$. For each of these, we measure the chlorophyll concentration variance. Finally, we estimate $Var(|L)$  by  averaging these variances across subsets of the same length. The spatial correlation function is derived by means of the Auto Correlation Function using the Python package \texttt{statsmodels} (supplementary text).

To evaluate the goodness of our theoretical prediction (Eq.~\ref{eq:results14}) for the scaling of $Var(n|L)$, we compare it with a power-law relation using a $\Delta_{\textrm{BIC}}$ test (supplementary text). As reported in Fig.~\ref{fig:SM_patch}, for chlorophyll concentrations above 0.1~$\textrm{mg}\,\,\textrm{m}^{-3}$ our model outperforms the simple power-law fit, whereas it does not reproduce the empirical patchiness for lower concentrations (oligotrophic regions). In the supplementary text, we replicate the correlation analysis presented in panel~C of Fig.~\ref{fig:3} after restricting the dataset to non-oligotrophic samples, thereby demonstrating improved agreement between empirical patterns and model predictions.


\subsection*{Supplementary Text}

\subsubsection*{Two-point correlation function}\label{Methods:2point}
In this section, we outline the key steps involved in deriving the population patterns discussed in the main text. The principal results are summarized in Fig.~\ref{fig:Methods}. Firstly, the core of the spatial analysis is the correlation function (Eq.~\ref{eq:Correlation}). To analyze the properties of such a function, we find the dynamical evolution of $\langle n_xn_y\rangle$ from Eq.~\ref{eq:model_intro}, which reads
\begin{equation}\label{eq:SM_2p_1}
    \partial_t\langle n_xn_y\rangle= 2\beta \langle n\rangle +2 g \langle n_x n_y\rangle - \frac{U(x,y)}{c}-2\frac{\langle n\rangle\langle n^2\rangle}{c}+D\left(\partial_x^2+\partial_y^2\right)\langle n_xn_y\rangle +\sigma^2 \langle n^2\rangle \delta(x-y)
\end{equation}
where $U(x,y)=\langle n_x^2n_y\rangle+\langle n_y^2n_x\rangle-2\langle n\rangle\langle n^2\rangle$, and $\delta(x-y)$ is the Dirac delta function. $U$ incorporates the non-linearity of the system resulting from the quadratic death rate. Next, we observe that, from the definition of the two-point correlation function, we have $\langle n_x n_y\rangle=C(x,y)+\langle n\rangle^2$. Consequently, in the stationary limit we can rewrite Eq.~\ref{eq:SM_2p_1} as follows:
\begin{equation}\label{eq:SM_2p_2}
    2g C(x,y)-\frac{U(x,y)}{c}+ D\left(\partial_x^2+\partial_y^2\right)C(x,y)+\sigma^2\langle n^2\rangle \delta(x-y)=0
\end{equation}
which corresponds to Eq.~\ref{eq:results5} in the Methods section.  Finally, eliminating the site notation $(x,y)$  and rewriting Eq.~\ref{eq:SM_2p_2} in terms of $r=|x-y|$ only, we arrive at the following equation
\begin{equation}\label{eq:methods1}
    2g C(r)-\frac{U(r)}{c}+2 D\partial_r^2 C(r)+\sigma^2 \langle n^2\rangle \delta(r)=0
\end{equation}
In the next step, we treat $U(x,y)=U(r)$ as a function of  $C(r)$. Indeed,  in the small noise regime, we have $U(r) \approx 2\langle n\rangle C(r)$. We posit that this linear relation is generalizable to arbitrary noise. With this ansatz, we can capture the $r$-dependence of $U(r)$ in terms of the $r$-dependence of the corresponding correlation function.  Hence, as discussed in~\cite{methods}, we approximate $U(r)=A+B\,C(r)$. For $r\to0$, we have $2(\langle n^3\rangle -\langle n\rangle\langle n^2\rangle )=A+B\,Var(n)$, while for $r \to \infty$ we find $A=0$. Therefore, the constant $B$ is given by:
\begin{equation}
    B=2\left(\frac{\langle n^3\rangle - \langle n^2\rangle\langle n\rangle}{Var(n)}\right)
\end{equation}
Putting the linear ansatz in Eq.~\ref{eq:methods1}, we obtain
\begin{equation}\label{eq:SM_2p_3}
    0 = g C(r)-\frac{B}{2 c}C(r)+D\partial_r^2C(r)+\frac{\sigma^2}{2}\langle n^2\rangle \delta(r)
\end{equation}
From the last equation, we recognize the expression for the characteristic time (see~\cite{Peruzzo2020_neutral_space}): 
\begin{equation}\label{eq:SM_2p_4}
    \tau^{-1}= \frac{B}{2 c}-g
\end{equation} 
At this point, we solve Eq.~\ref{eq:SM_2p_3} using Fourier transformation:
\begin{equation}
    \hat{C}(k)=\int_{\mathbf{R}}dr\,C(r)\,e^{-\mathrm{i} r k}
\end{equation}
Applying this transformation, Eq.~\ref{eq:SM_2p_3} becomes
\begin{equation}
    \frac{\hat{C}(k)}{\tau}+D k^2\hat{C}{(k)}=\frac{\sigma^2}{2}\langle n^2\rangle
\end{equation}
and
\begin{equation}
    \hat{C}(k)=\frac{\tau\sigma^2\langle n^2\rangle}{2\left(1+k^2\Gamma^2\right)}
\end{equation}
where we have $\Gamma=\sqrt{\tau D}$. Using the inverse transformation, we obtain the following:
\begin{equation}
    C(r)=\frac{1}{2\pi}\int dk\,\hat{C}(k)\,e^{\mathbf{i}kr}=\frac{\tau \sigma^2\langle n^2\rangle}{4\pi }\int dk\,\frac{e^{\mathbf{i}kr}}{1+k^2\Gamma^2}
\end{equation}
The formal solution of this equation is
\begin{equation}\label{eq:SM_2p_4}
    C(r)=\frac{\tau\sigma^2\langle n^2\rangle}{4\Gamma}e^{-|r|/\Gamma}
\end{equation}
which is exactly the exponential two-point correlation function reported in the main text. In panel~A of Fig.~\ref{fig:Methods} we show a comparison between our theoretical prediction and numerical simulations of Eq.~\ref{eq:model_intro}.

\subsubsection*{Characteristic spatiotemporal scales}\label{Methods:spatio_temporal_scales}
Here we discuss the derivation of $\tau$ and $\Gamma$.  As shown in~\cite{methods}, the characteristic time $\tau$ can be rewritten in terms of the moments of $n$, which we can estimate starting from the effective Langevin equation (Eq.~\ref{eq:results10}). Indeed, following the discussion in the main text and in~\cite{methods}, we observe that Eq.~\ref{eq:results10} reproduces the stationary mean and variance of the complete system (Eq.~\ref{eq:model_intro}). Let us now assume that the effective dynamics also approximates the third moment $\langle n^3\rangle$. In doing so, we are able to express $\tau$ as a function of the parameters of the effective GIG, since:
\begin{equation}\label{eq:SM_moments}
    \langle n\rangle=\frac{ \sqrt{\beta c } K_{\frac{2 g }{{\sigma^\star} ^2}}\left(\rho\right)}{K_{\frac{2 g }{{\sigma^\star}^2}-1}\left(\rho\right)}\,,\qquad \langle n^2\rangle =\frac{\beta  c K_{\frac{2 g }{{\sigma^\star} ^2}+1}\left(\rho\right)}{K_{\frac{2 g }{{\sigma^\star} ^2}-1}\left(\rho\right)}\,,\qquad \langle n^3\rangle = \frac{(\beta  c)^{3/2} K_{\frac{2 g }{{\sigma^\star} ^2}+2}\left(\rho\right)}{K_{\frac{2 g }{{\sigma^\star} ^2}-1}\left(\rho\right)}
\end{equation}
where \(\rho=4\sqrt{\beta/c}{\sigma^\star}^{-2}\), and $\sigma^{\star}$ is defined in Eq.~\ref{eq:results10}. Therefore, using the expression for the power-law exponent $\lambda$ introduced in the main text,
\begin{equation}\label{SM:eq_lambda}
\lambda=2\left(1-\frac{g}{{\sigma^\star}^2}\right)
\end{equation}
we find that Eq.~\ref{eq:SM_2p_4} becomes:
\begin{equation}
    \tau=\frac{2}{{\sigma^\star}^2}\left[1 - \frac{K_{2-\lambda}(\rho)^2}{K_{\lambda-1}(\rho)\,K_{3-\lambda}(\rho)} \right]
\end{equation}
Recalling that $\Gamma=\sqrt{D \tau}$, we can rewrite the previous equation as follows:
\begin{equation}\label{eq:SM_self_cons}
    \Gamma=\frac{4 D}{\sigma^2}\left[1-\frac{K_{2-\lambda(\Gamma)}(\rho(\Gamma))^2}{K_{\lambda(\Gamma)-1}(\rho(\Gamma))K_{3-\lambda(\Gamma)}(\rho(\Gamma))}\right]
\end{equation}
where we use the notation $\lambda(\Gamma)$ and $\rho(\Gamma)$ to highlight the dependency on $\Gamma$ given by the effective noise. Eq.~\ref{eq:SM_self_cons} is a self-consistent equation for $\Gamma$. Fig.~\ref{fig:Methods}-B shows a comparison between the numerical solution of this equation and the correlation length obtained from the numerical simulations of Eq.~\ref{eq:model_intro}. We observe a strong consistency, supporting the validity of our approach. 

Since the empirical values of $\lambda$ obtained in our analysis are clustered around $\lambda=1.5$, we solve Eq.~\ref{eq:SM_self_cons} for this specific slope to derive an approximation for the correlation length. We observe that the choice of this $\lambda$ leads to a simplification of Eq.~\ref{eq:SM_self_cons}, which becomes:
\begin{equation}
    \Gamma=\frac{4 D}{\sigma^2}\left[1-\left(1+\frac{\sqrt{c/\beta}\sigma^2}{8\Gamma}\right)^{-1}\right]
\end{equation}
whose solution is
\begin{equation}\label{eq:SMGamma_app1}
    \Gamma^{(1)}=\frac{\sqrt{c/\beta}\sigma^2}{16}\left(\sqrt{128 D \sigma^{-4} \sqrt{\beta/c}+1}-1\right)
\end{equation}
This approximation results valid for a wide range of parameters as proven in panel~A of Fig.~\ref{fig:SM_Gamma}. 
Since we generally expect $\beta\ll 1$ and $c\gg 1$, assuming relatively large values of $\sigma$, the previous equation can be approximated as follows:
\begin{equation}\label{eq:SMGamma_app2}
\Gamma^{(2)}\approx\frac{4 D}{\sigma^2}
\end{equation}
which corresponds to the approximation given in Eq.~\ref{eq:Gamma_approx} of the main text. The relative error of $\Gamma^{(2)}$, defined as $\epsilon=2\vert{}\Gamma^{(1)}-\Gamma^{(2)}\vert{}/(\Gamma^{(1)}+\Gamma^{(2)})$, is shown in panel~B of Fig.~\ref{fig:SM_Gamma}, confirming the accuracy of this approximation for $\beta/c\ll1$ and large $\sigma$

\subsubsection*{Spatial variance}\label{Methods:spatial_variance}
From $C(r)$, we immediately derive the variance of the total population of a transect $L$, $Var(n_L)$. We have to find an equation for $\langle n_L^2\rangle $, where $n_L$ is the total population along a transect of length $L$.  Let us now return to Eq.~\ref{eq:SM_2p_1} for the coupled moment $\langle n_x n_y\rangle$. Following~\cite{Peruzzo2020_neutral_space}, we integrate $\langle n_x n_y\rangle$ with respect to both $x$ and $y$ over $L$. In doing so, we obtain the  second moment of the total population:
\begin{equation}
    \langle n_L^2\rangle=\int_0^L dx\int_0^Ldy\, \langle n_xn_y\rangle 
\end{equation}
Using the expression for $C(x,y)$, this equation becomes
\begin{align}
    \langle n_L^2\rangle&=L^2\langle n\rangle^2+Var(n)\int_0^L dx\int_0^Ldy\,e^{\frac{-|x-y|}{\Gamma}} \\
    &=L^2\langle n\rangle^2+Var(n)\int_0^Ldx\left(\int_x^Ldy\, e^{\frac{-y+x}{\Gamma}}+\int_0^xdy\,e^{\frac{-x+y}{\Gamma}} \right)\\&=L^2\langle n\rangle^2 +\Gamma Var(n)\int_0^Ldx\,\left(2-e^{-\frac{x}{\Gamma}}-e^{\frac{x-L}{\Gamma}}\right)\\
    &=L^2\langle n\rangle^2+2\Gamma L Var(n) \left(1-\frac{2 \Gamma}{L}e^{-\frac{L}{2\Gamma}}\sinh{\left(\frac{L}{2\Gamma}\right)}\right)
\end{align}
Then, we recover the spatial variance reported in the main text
\begin{equation}\label{eq:SM_var}
Var(n_L)=2\Gamma L Var(n)\psi\left(\frac{L}{2 \Gamma}\right)
\end{equation}
where we define
\begin{equation}\label{eq:SM_psi}
    \psi(z)=1-e^{-z}\sinh{(z)}/z
\end{equation}
with $z>0$. A validation of this theoretical prediction with the spatial variance evaluated simulating Eq.~\ref{eq:model_intro} is presented in Fig.~\ref{fig:Methods}-C. Note that when $L=u$, since generally $L\gg u$ we have $Var(n_u)\approx u^2 Var(n)$, which is the variance employed for the derivation of Eq.~\ref{eq:results10}. From this spatial variance, we immediately derive the scaling of the patchiness. In fact, as shown in Methods~\ref{subsec:patchiness}, patchiness is defined by the conditional variance of a transect:
\begin{equation}\label{eq:sub_pat_1SM}
     Var(n|L)=\left\langle \frac{1}{L}\int_{L}dx \left(n_x -\int_Ldy\, n_y/L\right)^2 \right\rangle
\end{equation}
where $L^{-1}\int_L n_xdx$ is the effective mean of $n_x$ in a sample of size $L$. Therefore, as $L$ becomes large, this mean converges to the expectation value $\langle n\rangle$ when the field is ergodic.  By expanding the quadratic term in the last equation and employing the definition of total abundance $n_L$,   the expectation value of the empirical variance becomes
\begin{equation}
    Var(n|L)=Var(n)-\frac{Var\left(n_L\right)}{L^2}
\end{equation}
Using the scaling law~\ref{eq:results13}, we recover Eq.~\ref{eq:results14}. This expression is compared with simulated samples in Fig.~\ref{fig:Methods}-D. 

\subsubsection*{Local distribution and Fokker-Plank equation}\label{SM:local_distribution}
The probability distribution reported in Eq.~\ref{eq:local_dist} is the stationary solution of the so-called Fokker-Planck equation~\cite{gardiner2009_stochastic} associated to the Langevin dynamics in Eq.~\ref{eq:model_intro}. We rewrite the stochastic differential equation as $\partial_t n= n\bar{\mu}(n)+\sqrt{\bar\Sigma(n)}\xi_t$, with $\bar\Sigma(n)$ the noise amplitude such that $\Sigma=\langle \bar\Sigma\rangle$ (see Eq.~\ref{eq:Sigma}). With this notation, the Fokker-Planck equation is:
\begin{equation}
    \partial_tP(n)=-\partial_nJ(n)
\end{equation}
where we introduce the probability flux
\begin{equation}
    J(n)=n\bar{\mu}(n)P(n)-\partial_n \left[\frac{\bar\Sigma(n)}{2}P(n)\right]
\end{equation}
At the stationary limit, we have $\partial_t P(n)=0$. Moreover, we impose reflecting boundary conditions, namely the flux of probability vanishes at $n=0$. Therefore, the stationary distribution is the solution of $J(n)=0$, which reads
\begin{equation}
    P(n)=\mathcal{N} 2\bar\Sigma^{-1}(n)\exp{\left(\int^n dm\,\frac{2 n\bar{\mu} (m)}{\bar \Sigma(m)}\right)}
\end{equation}
with $\mathcal{N}$ the normalization factor. Using the explicit expressions for $\bar{\mu}$ and $\bar \Sigma(n)$, we find the GIG for the case without diffusivity:
\begin{equation}
    P(n)=\frac{ \left( \beta c \right)^{\tfrac{\lambda-1}{2}} 
\, e^{-\tfrac{2}{{\sigma}^2} \left(\tfrac{\beta}{n} + \tfrac{n}{c}\right)} 
\, n^{-\lambda} }
{ 2 \, K_{\lambda-1} \!\left( \tfrac{4 \sqrt{\beta/c}}{{\sigma}^2} \right) }
\end{equation}
where the normalization factor is chosen to have $\int_0^\infty dn\,P(n)=1$.

\subsubsection*{Different regimes of the probability distribution}\label{SM:transition}

In this section, we analyze the regime transition induced by diffusion. As commented in the main text, the competition between diffusion $D$ and local noise $\sigma$ can be described by two regimes: when diffusion becomes dominant over local noise, we have relatively larger correlation lengths, and the homogenizing effects become large enough to influence the statistical properties of the population. In this regime, we expect $P(n)$ to be well approximated by a log-normal distribution. On the other hand, when $D$ becomes smaller relative to local fluctuations, we expect $P(n)$ to show a power-law regime and find a GIG distribution, as in the case without diffusion. 
To make these two regimes explicit, we recall the change of variables introduced in~\cite{methods}, $Y_x = \log(n_x/ \langle n \rangle)$. This transformation allows us to focus the analysis on the behavior of the growth rate, rather than on the population itself. Using Itô's lemma~\cite{gardiner2009_stochastic}, the stochastic equation for $Y_x$ can be derived directly from Eq.~\ref{eq:model_intro}:
\begin{equation}
\partial_tY_x=g+\frac{\beta}{\langle n\rangle}e^{-Y_x} -\frac{\langle n\rangle }{c}e^{Y_x}-\frac{\sigma^2}{2}+D(\partial_x^2Y_x+(\partial_xY_x)^2)+\sigma \xi_x(t)
\end{equation}
We emphasize that all calculations are defined on a discrete lattice of spacing $u$. Although this equation has a more complicated drift term than Eq.~\ref{eq:model_intro}, it features purely additive noise. 

We assume that the distribution of $Y_x$ is peaked at zero, that is, $n_x\sim\langle n\rangle$, and expand the latter equation as follows:
\begin{equation}\label{eq:SM_trans1}
\partial_tY_x=\frac{\beta+g \langle n\rangle -\langle n\rangle^2 c^{-1}-\sigma^2\langle n\rangle /2}{\langle n\rangle }+\sum_{j\ge 1}I_jY_x^j+
    D\left(\partial^2_xY_x+(\partial_xY_x)^2\right)+\sigma \xi_x(t)
\end{equation}
with
\begin{equation}\label{eq:SM_trans2}
    I_j=\frac{\beta (-1)^j-c^{-1}\langle n\rangle^2}{j!\langle n\rangle}
\end{equation}
We note that, near stationarity, the first term of Eq.~\ref{eq:SM_trans1} can be rewritten as:
\begin{equation}\label{eq:SM_trans3}
\frac{\beta+g \langle n\rangle -\langle n\rangle^2 c^{-1}-\sigma^2\langle n\rangle /2}{\langle n\rangle }=\frac{Var(n)}{c\langle n\rangle}-\frac{\sigma^2}{2}
\end{equation}
where we used the stationary relation $\beta+g\langle n\rangle -c^{-1}\langle n^2\rangle=0$. For small fluctuations, this term can be neglected. In any case, reintroducing these constant terms would merely result in a shift of $\langle Y_x\rangle$. In doing so, we recognize the characteristic time scale of $Y_x$, $\tau_{y}=-I_1^{-1}$ and the characteristic correlation length, $\Gamma_Y=\sqrt{D\tau_Y}$. Then, analogously to the study of $n$, we expect to find that the amplitude of the local fluctuations of $Y_x$ scales as $\tilde{\sigma}=\sigma (2 \Gamma)^{-1/2}$, that is,  damped by spatial diffusivity. Therefore, $Var(Y_x)\propto \epsilon^2$ with $\epsilon=\tau_Y^{1/2}\tilde{\sigma}$. 
For large $\Gamma$, we have $\epsilon\ll1$, and we employ this quantity as a small parameter to expand $Y_x$:
\begin{equation}\label{eq:T_espilon_exp}
    Y_x=\sum_i\epsilon^i Y_x^{(i)}
\end{equation}
To show the two regimes for $n$, we first analyze the weight of dispersal in terms of $\epsilon$. Then, we derive a criterion that justifies truncating the expansion at first order of $\epsilon$, thereby yielding a lognormal distribution for $P(n_x)$, and we demonstrate that this criterion is exactly equivalent to the condition $\lambda < 1.5$.

We observe that it is convenient to apply the following re-scaling: $t = T\,\tau_{Y}$ and $x = X\,\Gamma$. In doing so, we arrive at:
\begin{equation}
    \partial_T Y_X=-Y_X+I_2 \tau_{Y} Y_X^2 + \partial_X^2 Y_X+(\partial_X Y_X)^2+\sqrt{2}\tilde{\sigma}\xi_X(T)+\mathcal{O}(Y_X^3)
\end{equation}
By putting Eq.~\ref{eq:T_espilon_exp} into the previous equation, we find at order $\epsilon$ that
\begin{equation}\label{eq:sm_trnas}
     \partial_TY_X^{(1)}= - Y^{(1)}_{X}+ \partial_{X}^2 Y^{(1)}_X+\sqrt{2}\tilde{\sigma}\xi_{X}(T) 
\end{equation}
which is the spatial generalization of the Ornstein–Uhlenbeck processes. Because the solution is a linear functional of Gaussian noise, $Y_x(t)$ is a Gaussian process, fully specified by its mean and covariance. Finally,  by taking $Y_X\approx \epsilon Y^{(1)}_X$, we recover that $n_x$ is lognormal. 

\begin{figure}
    \centering
    \includegraphics[width=0.9\textwidth]{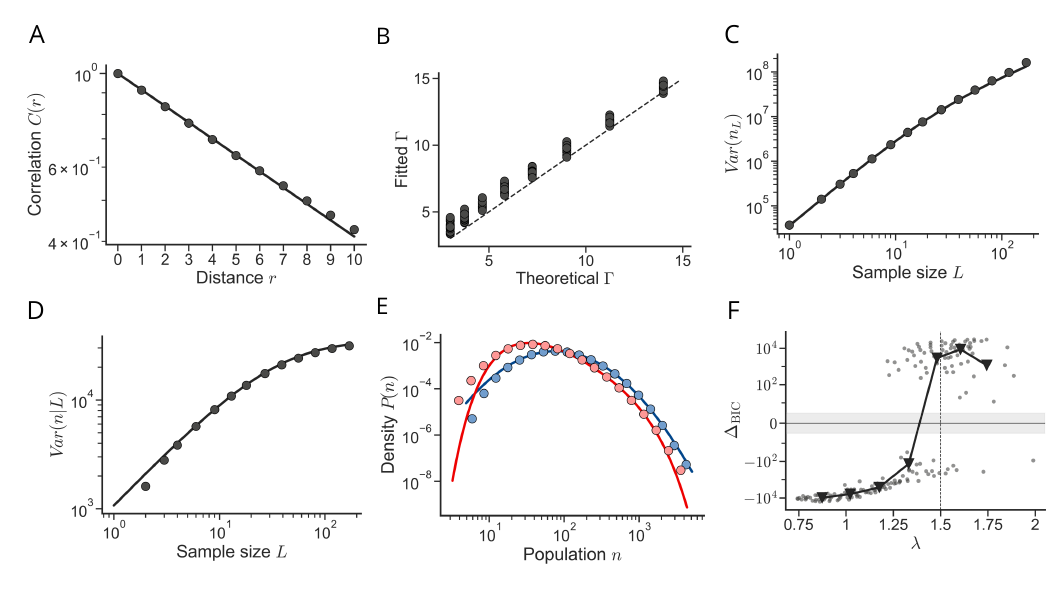}
    \caption{\textbf{Simulations and analytical predictions.} Numerical simulations of Eq.~\ref{eq:model_intro} are run on top of a one-dimensional lattice with $L=170$ sites, $u=1$ and periodic boundary conditions. We fix $\beta=0.3$, $c=10^{5}$ and $D=1$, while the other parameters are chosen to explore different values of $\lambda$. (\textbf{A}): we show the two-point correlation function with $g=2.3\times10^{-3}$ and $\sigma=0.45$. The predicted correlation length is $\Gamma=11.2$ . The solid line represents the theoretical curve given by Eq.~\ref{eq:Correlation}, while dots results from the simulation. (\textbf{B}): comparison between the fitted correlation length and the theoretical one for 140 simulations performed for different values of $g$ and $\sigma$. These combinations are selected to have 20 different values of $\lambda$  between 1 and 2, and 7 values of $\Gamma$ spanning the range from 3 to 14. (\textbf{C}~and~(\textbf{D})): spatial Taylor's law (Eq.~\ref{eq:results13}) and empirical variance (Eq.~\ref{eq:results14}) in log-log scale with $g=2.3\times10^{-3}$ and $\sigma=0.45$. Solid lines are the analytical prediction and dots represent the simulation outcomes.  (\textbf{E}): stationary abundance distribution for the two population regimes. In red, we report the log-normal regime for which $\lambda=1.15$ and $\Delta_{\textrm{BIC}}<0$. The parameters are $g=3.5\times 10^{-3}$ and $\sigma=0.43$, and the correlation length is $\Gamma=11.2$. Red dots are the simulated data, while the curve follows Eq.~\ref{eq:Methods_P}. In blue, we show the GIG regime with $\lambda=1.84$ and $\Delta_{\textrm{BIC}}>0$. For this case, we choose $g=7.4\times 10^{-4}$ and $\sigma=0.46$. The value of $\Gamma$ is identical to that used in the log-normal regime. Here as well, the continuous line denotes the theoretical prediction, whereas the dots indicate the outcomes of the numerical simulations. \textit{Panel~F}: $\Delta_{\textrm{BIC}}$ as a function of the fitted slope $\lambda$ over 140 simulations. Triangles are median values of $\Delta_{\textrm{BIC}}$, and the gray band is the neutral interval $|\Delta_{\textrm{BIC}}|<5$.}
    \label{fig:Methods}
\end{figure}

\begin{figure}
    \centering
    \includegraphics[width=0.95\linewidth]{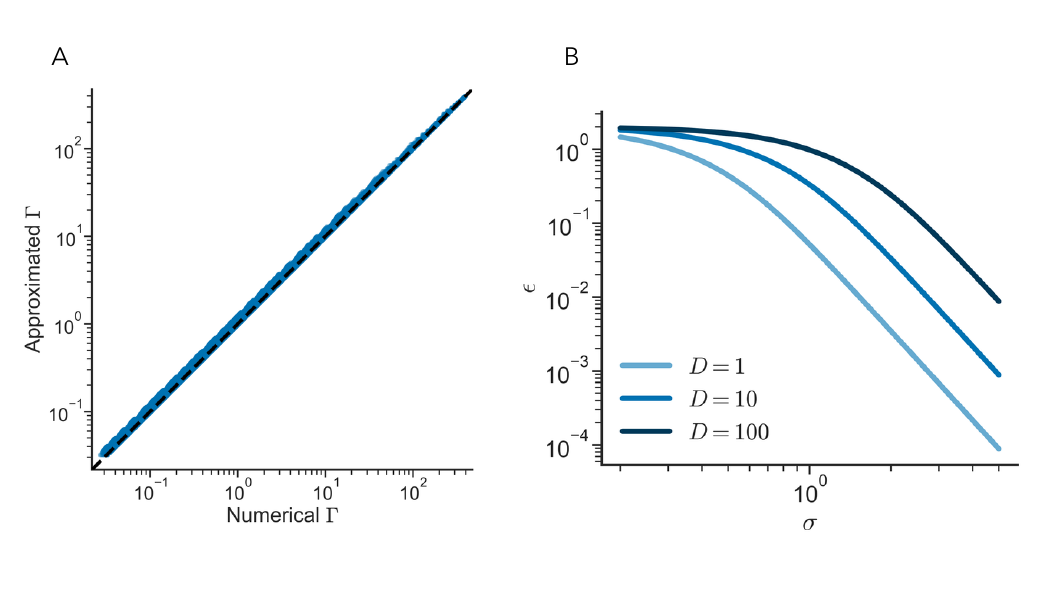}
    \caption{\textbf{Analytical expression for $\Gamma$.} (\textbf{A}): comparison between the numerical solution of Eq.~\ref{eq:SM_self_cons} (horizontal axis) and the approximation given by Eq.~\ref{eq:SMGamma_app1}. Dots denote the results of numerical evaluation of $\Gamma$ carried out for $4\times 10^4$ parameter combinations, with $\beta=0.3$ and $c=10^5$ held fixed, $D\in[0.1,160]$, and $\sigma$ selected to ensure that $\lambda\in[1,2]$. (\textbf{B}): relative error $\epsilon$ between $\Gamma^{(1)}$ and $\Gamma^{(2)}$ as a function of $\sigma$. We set $\beta=0.1$, $c=10^5$ and $D=1,\,10,\,100$.  }
    \label{fig:SM_Gamma}
\end{figure}

\begin{figure}
    \centering
    \includegraphics[width=1\linewidth]{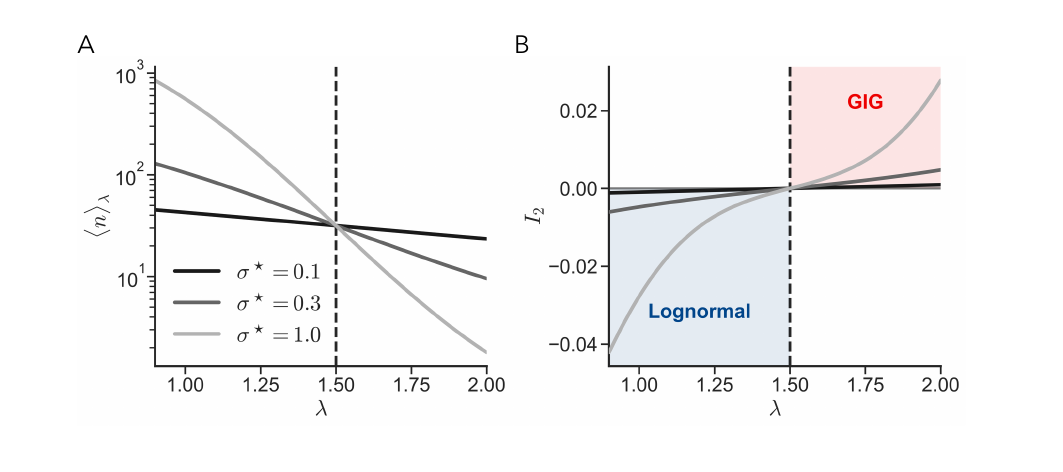}
    \caption{\textbf{Regime transition.}  (\textbf{A}):  Plot of $\langle n\rangle_\lambda$ (Eq.~\ref{eq:SM_meanlambda}) as a function of $\lambda$. Different colors indicates different noise amplitudes $\sigma^\star$. We consider $\beta=0.1$, $c=10^4$, and $g={\sigma^\star}^2(2-\lambda)/2$.  (\textbf{B}): Plot of $I_2$ (Eq.~\ref{eq:SM_I}) with respect to $\lambda$. For $\lambda<1.5$, $I_2$ is always negative while it becomes positive when $\lambda>1.5$.  }
    \label{fig:SM_transition}
\end{figure}

Now, let us show when we can neglect these $\mathcal{O}(\epsilon^2)$ terms. First, we recall the expressions for $\langle n\rangle$ (Eq.~\ref{eq:SM_moments}) and for $\lambda$ (Eq.~\ref{SM:eq_lambda}), and rewrite the population mean as a function of the power-law slope:
\begin{equation}\label{eq:SM_meanlambda}
    \langle n\rangle_{\lambda}=\frac{ \sqrt{\beta c } K_{2-\lambda}\left(\rho\right)}{K_{1-\lambda}\left(\rho\right)},
\end{equation}
with \(\rho=4\sqrt{\beta/c}{\sigma^\star}^{-2}\).
Substituting this expression into Eq.~\ref{eq:SM_trans2}, the coefficients $I_j$ become:
\begin{equation}\label{eq:SM_I}
I_j=\frac{(-1)^j\langle n\rangle_{3/2}^2-\langle n\rangle_\lambda^2}{j!,c,\langle n\rangle_\lambda} , .
\end{equation}
As shown in Fig.~\ref{fig:SM_transition}-A, $\langle n\rangle_\lambda$ is a monotonically decreasing function of $\lambda$. Therefore, for $\lambda<3/2$ we have $I_j<0$ for any $j$, while for $\lambda>3/2$ the even-order terms become positive (see panel~B in Fig.~\ref{fig:SM_transition}). In this latter case, truncating the expansion at the first order is no longer a valid representation of the system. Instead, one must retain all terms in the series to balance the even and odd contributions. By considering the full expansion, the dynamics are better captured by a GIG distribution. On the other hand, for $\lambda<3/2$ we can rely on Eq.~\ref{eq:sm_trnas} recovering the lognormal distribution. This argument provides a phenomenological interpretation of the regime shift at $\lambda=3/2$. Consequently, $P(n)$ can be written as follows:
\begin{equation}\label{eq:Methods_P} 
                    P(n) \propto 
                    \begin{cases}
                        \frac{1}{n} e^{-\frac{(\ln n - m)^2}{2\sigma_l^2}} & (\lambda < 3/2) \\[4pt]  
                        n^{-\lambda} e^{-\frac{2}{{\sigma^\star}^2} \left(\frac{\beta}{n} + \frac{n}{c}\right)} & (\lambda > 3/2)
                    \end{cases} 
\end{equation}
where the parameters of the lognormal regime are set to reproduce the population mean and variance of Eq.~\ref{eq:model_intro}, that is:
\begin{equation}
    \gamma=\frac{Var(n)}{\langle n\rangle^2} ,\qquad m=\ln\left(\frac{\langle n\rangle}{\sqrt{1+\gamma}}\right),\qquad \sigma_l=\sqrt{\ln(1+\gamma)}
\end{equation}
The analytical predictions of $P(n)$ are compared with the numerical simulations in panel~E of Fig.~\ref{fig:Methods}. We find excellent agreement for large abundances, while there are some deviations at $n<10$. Indeed, our analysis is based on the first two moments of $n$, which are dominated by the largest abundances. For small values of $n$, we expect that diffusion influences the system in a more complex way. The regime transition is confirmed in panel~F of Fig.~\ref{fig:Methods}, where we show $\Delta_{\textrm{BIC}}$ as a function of $\lambda$ for 140 simulations.

\subsubsection*{MareChiara: seasonality of SADs and TADs}
In the main text, the construction of both SADs and TADs does not take seasonality into account, but only captures the overall stochasticity of the population. 
As illustrated in panels~A~and~B of Fig.~\ref{fig:Methods_deltaBIC}, when examining the full time series, SADs appear to match our expectations with $\Delta_{\textrm{BIC}}\le0$ for $\lambda<1.5$, whereas the TADs do not exhibit the behavior predicted from the theory. In fact, diatoms and dinoflagellates result to follow GIG distributions with $\lambda<1.5$, while coccolithophores follow a lognormal distribution with $\lambda>1.5$. 

\begin{figure}
    \centering
    \includegraphics[width=0.75\linewidth]{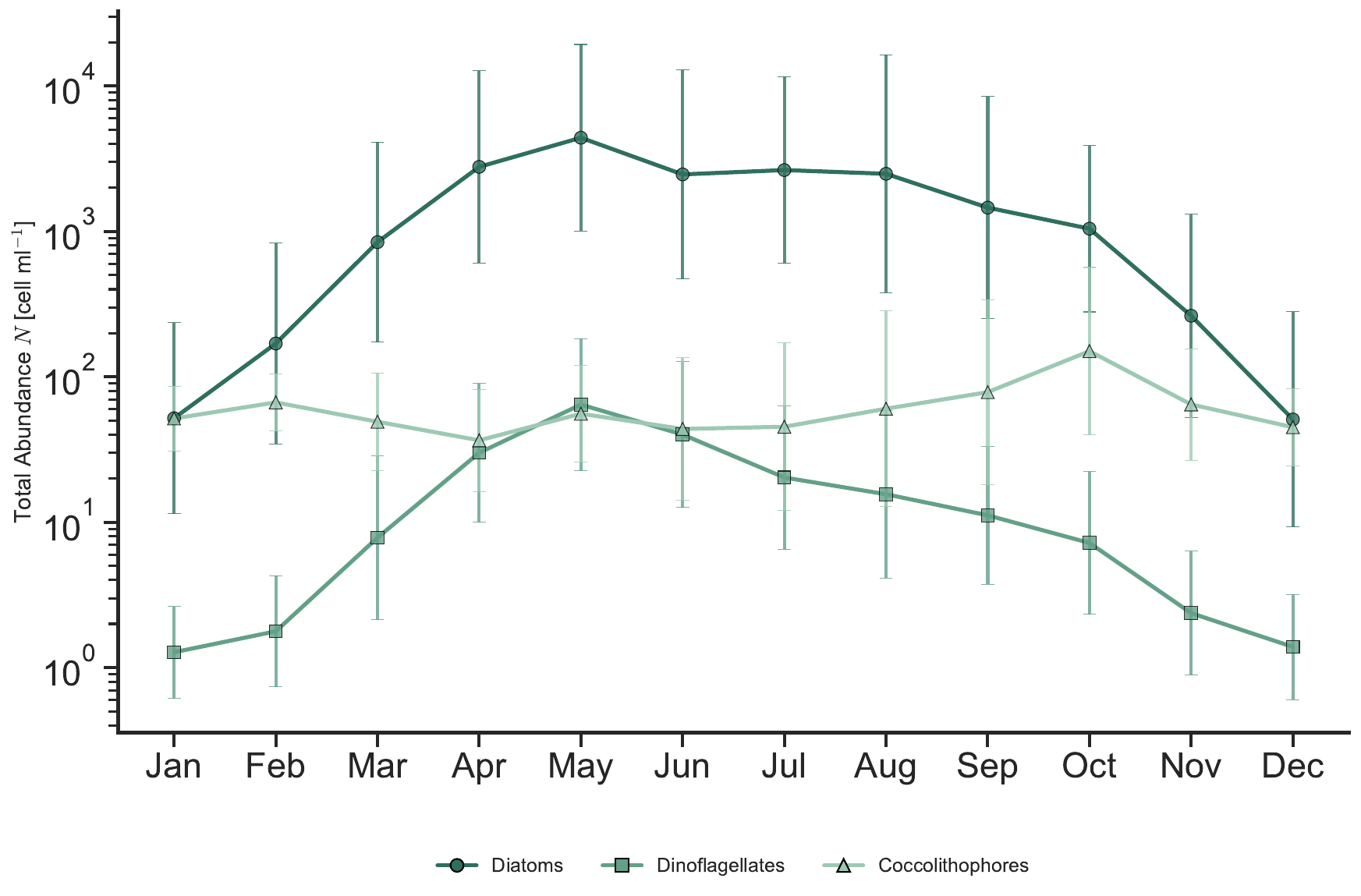}
    \caption{\textbf{Monthly trend of total abundances for MareChiara timeseries} Each point represents the geometric mean of the sample counts for the corresponding month across all years in the time series, while the error bars are based on the standard deviation of $\log(N)$. Different colors denote the three groups of plankton analyzed within this dataset.}
    \label{fig:Methods_seasonality_N}
\end{figure}

\begin{figure}
    \centering
    \includegraphics[width=1\linewidth]{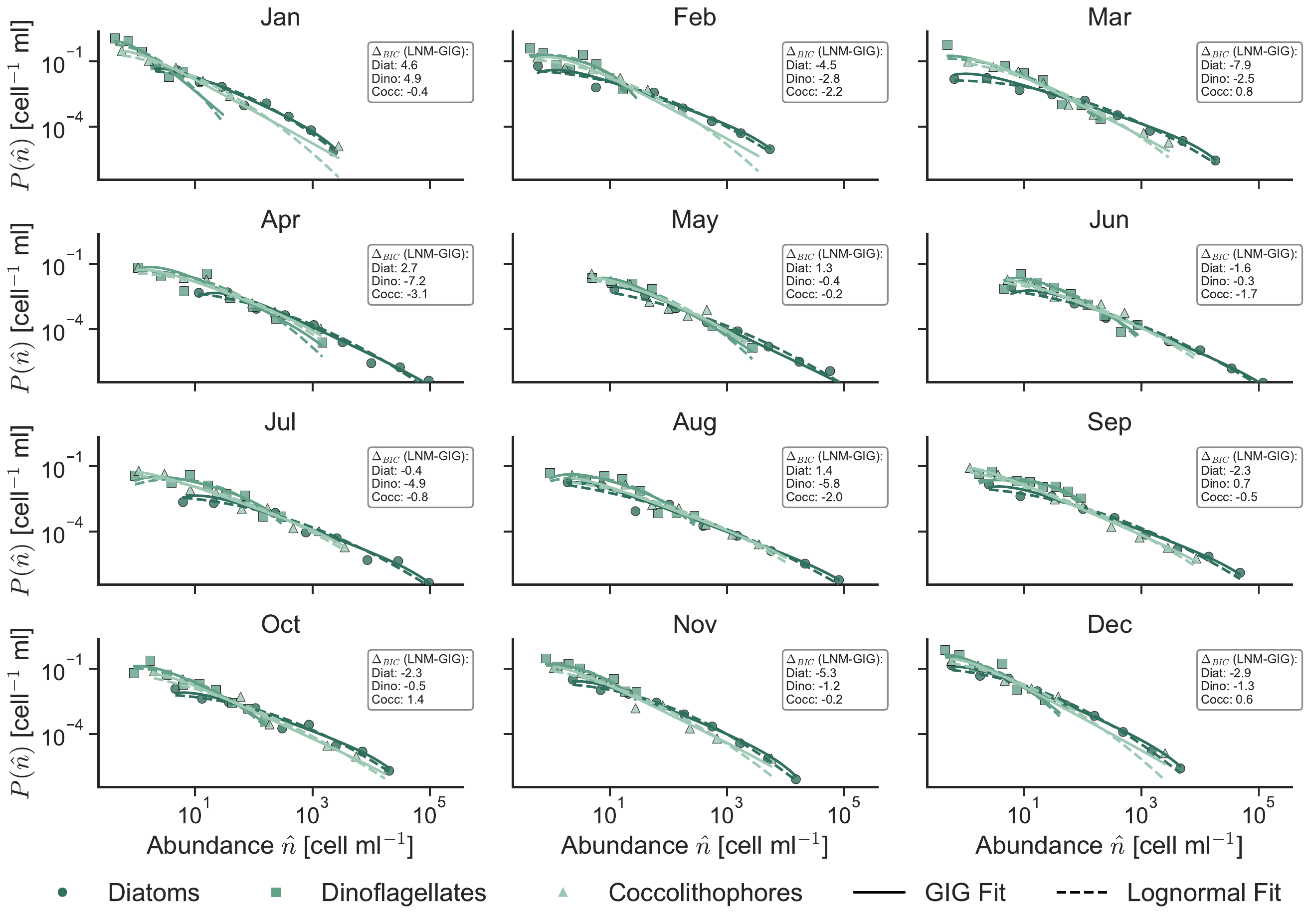}
    \caption{\textbf{Monthly SADs for MareChiara timeseries} Each panel represents the SADs obtained by summing, for each species within a single group, all sample counts from the same month over the entire duration of the time series. The distribution is derived by taking the normalized histogram of the total abundances among the species of a single group. The solid lines represent GIG fits, while the dashed lines represent fits of the log-normal distributions (see Eq.~\ref{eq:Methods_P}). The box in each panel shows the $\Delta_{\textrm{BIC}}$ values defined as in Methods~\ref{Methods:BIC}. The analysis is repeated for the three groups analyzed in the time series: diatoms, dinoflagellates, and coccolithophores. }
    \label{fig:Methods_SAD_MC}
\end{figure}

\begin{figure}
    \centering
    \includegraphics[width=1\linewidth]{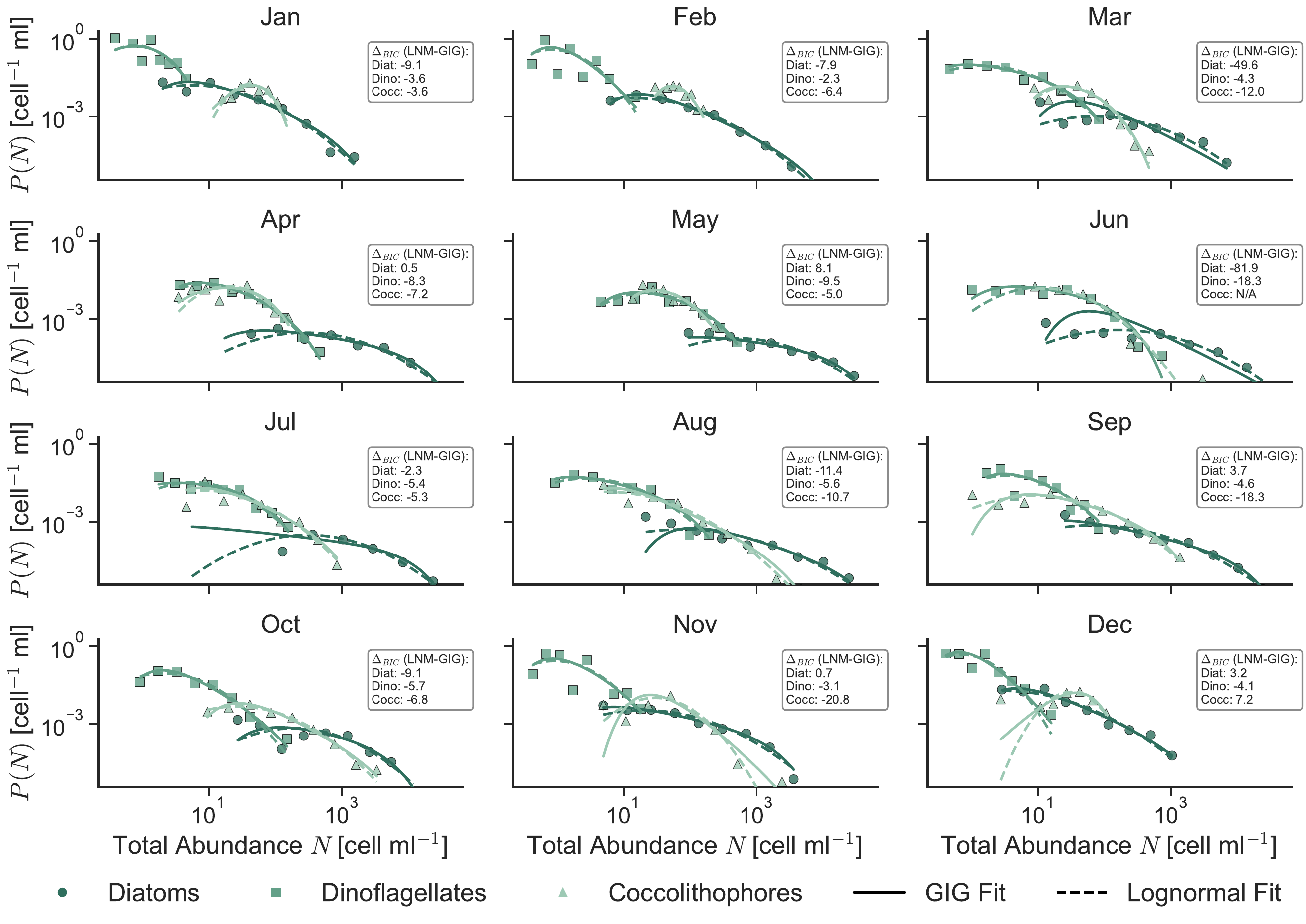}
    \caption{\textbf{Monthly TADs for LTER-MC dataset.} For each panel, we report the monthly TAD. For every group and every month, we compute, for each sample, the total counts summed across all species. The TAD is estimated using the normalized histogram of the total monthly counts across all years. The solid lines represent the fit with the GIG, while the dashed line is the fit with a log-normal distribution. For each plot we report the $\Delta_{\textrm{BIC}}$ values.  }
    \label{fig:Methods_TAD_MC}
\end{figure}

\begin{figure}
    \centering
    \includegraphics[width=0.85\linewidth]{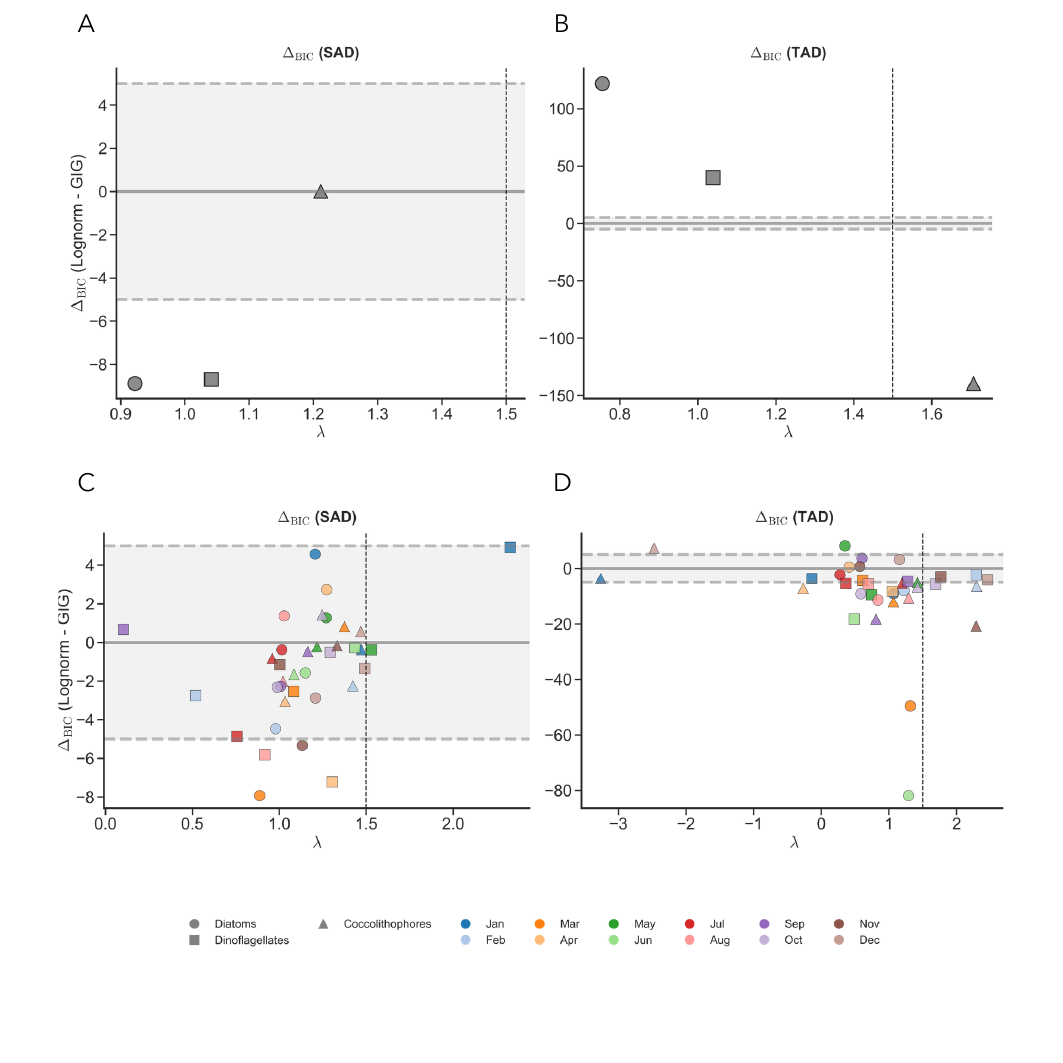}
    \caption{\textbf{$\Delta_{\textrm{BIC}}$ trend for LTER-MC dataset.} The vertical axis represents the $\Delta_{\textrm{BIC}}$ values in function of the fitted $\lambda$ values estimated for each month and each group. The gray region of $|\Delta_{\textrm{BIC}}|<5$ represents the range within which the two distributions are considered equivalent. (\textbf{A}~and~\textbf{B}): we report the trend for aggregated data without considering seasonality. In panel~A we find that the three groups have a lognormal-like SAD with $\lambda<1.5$, in accordance with our theory. Instead, panel~B displays the TAD’s $\Delta_\textrm{BIC}$ trend, which exhibits the opposite behavior to that predicted by our framework. (\textbf{C}~and~\textbf{D}): $\Delta_\textrm{BIC}$ trend for each month. Panel~C represents the trend for the SADs, while panel~D shows the TADs result. In both cases, we see that most of the samples with negative $\Delta_{\textrm{BIC}}$ have $\lambda<1.5$, which is consistent with our theoretical expectations.}
    \label{fig:Methods_deltaBIC}
\end{figure}

We now show that the inclusion of seasonality resolves this inconsistency. First, to visualize seasonal effects, we estimate the geometric mean value of the total counts across diatoms, dinoflagellates and coccolithophores for each month. As shown in Fig.~\ref{fig:Methods_seasonality_N}, diatoms and dinoflagellates exhibit a very similar trend, with a peak in abundance between April and May, while coccolithophores show a relatively more constant abundance with higher values around October. In all cases, we identify a monthly trend which cannot be described as an artifact of stochastic noise, but rather indicates a temporal change in the rates presented in Eq.~\ref{eq:model_intro}. Such variation can have multiple drivers, but the abundance profile seems to mostly reflect solar radiation.
Here, we repeat the SAD and TAD analysis described in the main text for each month separately. In Fig.~\ref{fig:Methods_SAD_MC} we show the monthly SADs. These plots are realized by taking the distribution of counts among the abundances of each species, summed across all years, while only considering samples taken in a certain month. Similarly, in Fig.~\ref{fig:Methods_TAD_MC} we present the various TADs. We note that the overall behavior of the distributions is similar to that discussed in the main text: SADs present longer tails compared to TADs, while the latter are, in general, more peaked around the median values.  The main differences compared to the plots analyzed in the main text are related to the position of the right cutoff, which reflects the seasonality of total abundance. By estimating the slope $\lambda$ for each month by fitting Eq.~\ref{eq:local_dist}, we obtain a monthly comparison between $\Delta_{\text{BIC}}$ and $\lambda$ for both focal species and total abundances. In panels~C~and~D of Fig.~\ref{fig:Methods_deltaBIC},  we find that the trend of $\Delta_{\text{BIC}}$ with respect to $\lambda$ is consistent with the transition at $\lambda= 1.5$, for both the SADs and the TADs. This suggests that the statistical properties of $\hat{n}$ and $N$ can be described by Eq.~\ref{eq:model_intro} by using monthly time windows, that is, by eliminating seasonal heterogeneity.

\subsubsection*{MareChiara: $\Sigma$ analysis}\label{SM:MC_noise}
In the main text (Fig. 2), we show the general scaling of $\Sigma$ for the abundance of the focal species $\hat n$ and for the total population $N$, using aggregated data without including seasonal effects. Here, we present the tests we conducted to confirm the environmental noise hypothesis for the aggregated data and the results of the analysis performed by separating the time series by months. 

\begin{figure}
    \centering
    \includegraphics[width=1\linewidth]{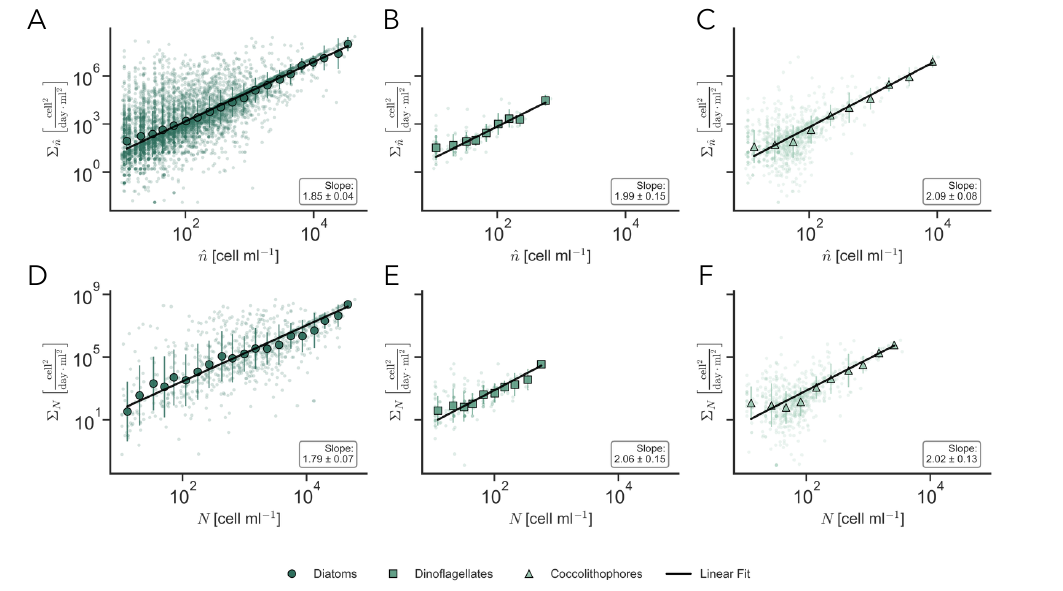}
    \caption{\textbf{Noise amplitude for focal species and total abundances from LTER-MC dataset.} For each bin of $\hat{n}$ and $N$, we show the corresponding values of $\delta n_T^2(n)/T$ (small dots), while large markers are their median values. (\textbf{A},~\textbf{B}~and~\textbf{C}): behavior of noise amplitude for focal species abundances. Panel~A reports the noise scaling for diatoms, panel~B for dinoflagellates while panel~C shows the noise amplitude for coccolithophores. (\textbf{D},~\textbf{E}~and~\textbf{F}): scaling analysis for total abundances. Panel~D illustrates the noise trend for diatoms, panel~E for dinoflagellates and panel~F for coccolithophores.  For both $\hat{n}$ and $N$, error bars are given by the standard deviation of $\log(\delta n_T^2/T)$, and black lines represent the linear interpolation.}
    \label{fig:Methods_S}
\end{figure}

First, in Fig.~\ref{fig:Methods_S} we show the trend of the geometric means of  $\delta n_T^2(\hat{n})/T$ for both $\hat{n}$ and $N$. We consider a threshold $n>10~$[cell ml$^{-1}$] for both focal and total abundances to analyze the noise scaling. Indeed, lower abundances are generally more noisy. In each panel of Fig.~\ref{fig:Methods_S}, we report the slope of the linear interpolation on the log-log scale, that is, 
\begin{equation}\label{eq:SM_Sigma_linear}
\log\left(\Sigma\right)=a+b\log(n)
\end{equation}
As shown, the slopes $b$ are consistent with the expected value 2 confirming the ansatz of a fluctuating per-capita growth rate.

\begin{figure}
    \centering
    \includegraphics[width=1\linewidth]{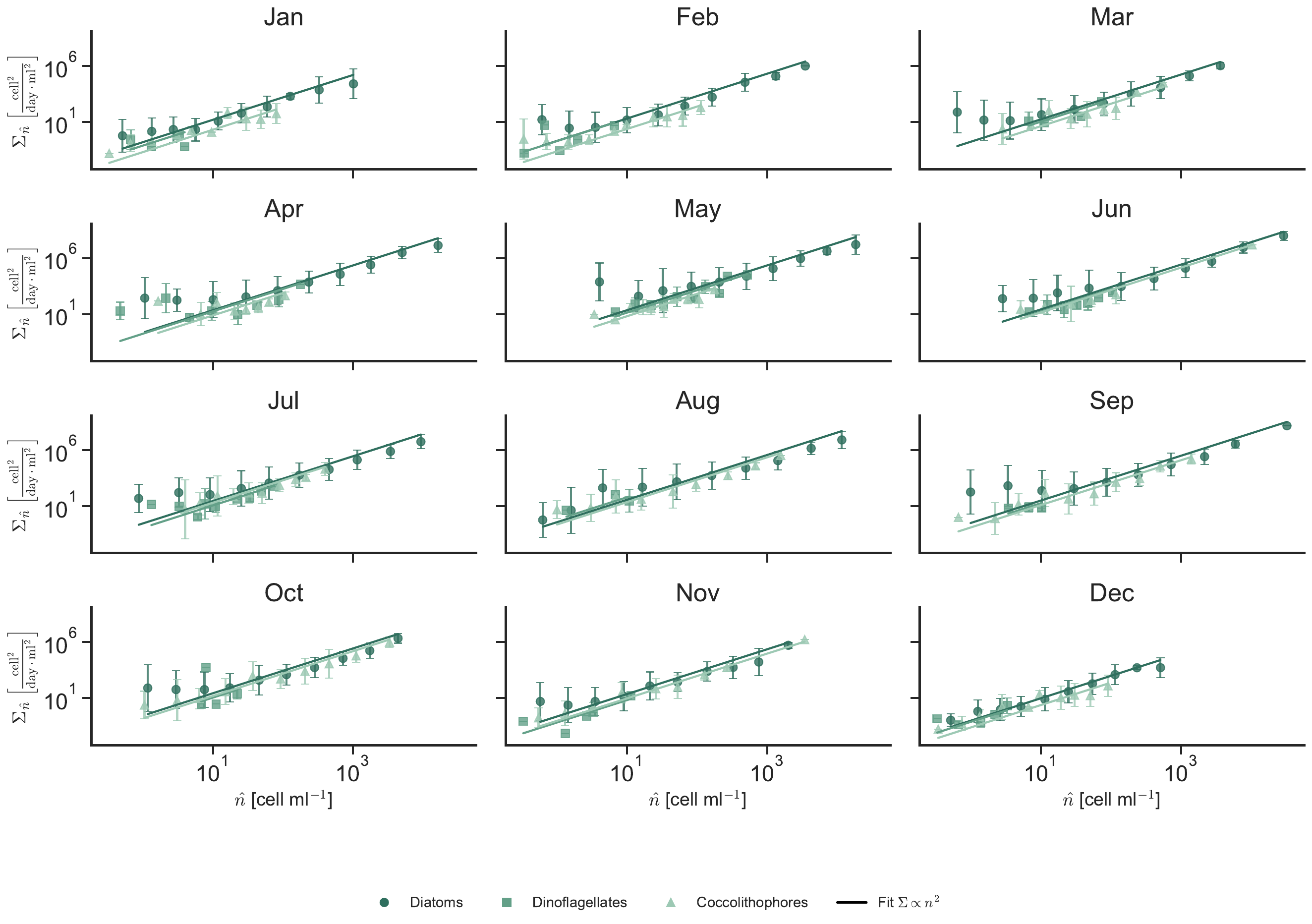}
    \caption{\textbf{Monthly noise amplitude of focal species from LTER-MC dataset.} Each panel shows the geometric means of $\delta n_T^2(n)/T$ for the abundances of individual species obtained using samples from a specific month along the whole time series. Error bars are given by the standard deviation of $\log(\delta n_T^2)$. Solid lines are the linear interpolation setting slope 2 on the log-log scale. }
    \label{fig:Methods_Sigma_n}
\end{figure}

\begin{figure}
    \centering
    \includegraphics[width=1\linewidth]{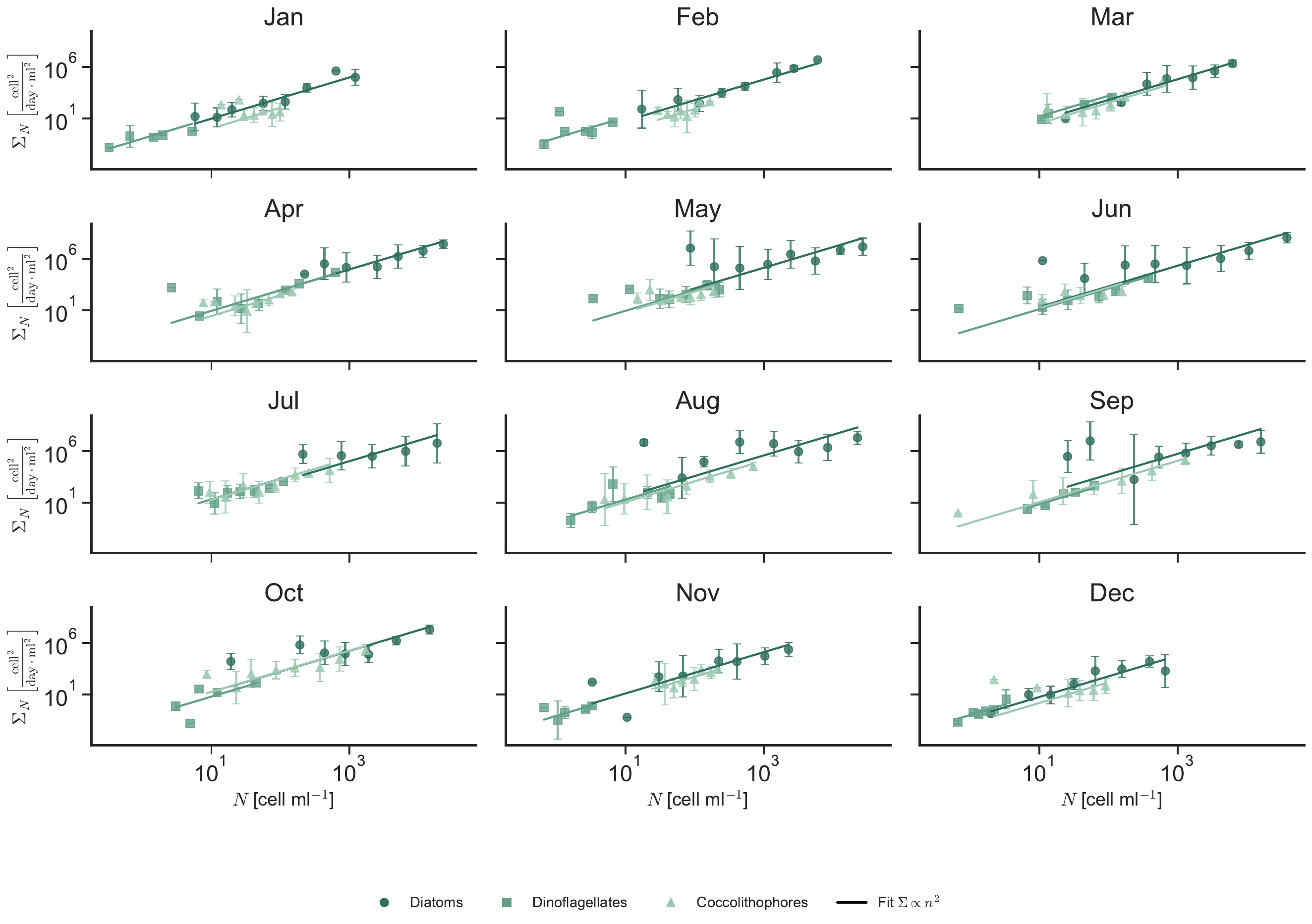}
    \caption{\textbf{Monthly noise amplitude of total abundance from LTER-MC dataset.} Each panel shows the geometric bin of  $\delta n_T^2/ T$  total abundance for each month. Error bars represent the standard deviation of $\log(\delta_T^2n/T)$, while the solid lines are the best of Eq.~\ref{eq:SM_Sigma_linear} setting $b=2$. }
    \label{fig:Methods_Sigma_N}
\end{figure}

Finally, in Figs.~\ref{fig:Methods_Sigma_n}~and~\ref{fig:Methods_Sigma_N}, we show the monthly scaling of $\Sigma$ for both $\hat{n}$ and $N$. As in the case of the aggregated data, we report the geometric means for different abundance bins. We emphasize that, given the reduced statistic compared to the entire time series, these trends should be considered as a qualitative behavior. However, for both focal species and total abundances, we clearly recognize the expected environmental scaling, providing further evidence of the applicability of our model on a monthly scale.

\subsubsection*{GRUMP: spatial correlations}\label{Methods:empirical_corr}

Here, we describe the analysis of the two point correlation function (Eq.~\ref{eq:Correlation}) using the GRUMP dataset. For each transect and sequence type (16S or 18S), we calculate the in-water distance $d_{ij}$ between each pair of the filtered samples using the R-packages \texttt{sf}~\cite{Pebesma2018_sf}, \texttt{gdistance}~\cite{vanEtten2017_gdistance} and \texttt{geosphere}~\cite{Hijmans2024_geosphere}. Then, we calculate the normalized ecological correlation between each pair of sampling stations ($i,j$) in the selected subset as \cite{Peruzzo2016_spatial_phenomenological}:
\begin{equation}
    \label{eq:PCF_emp}
    g_{ij} = \frac{\frac{1}{S}\sum_{\mu=1}^S n_{i\mu}n_{j\mu}}{\big(\frac{1}{S}\sum_{\mu=1}^S n_{i\mu}\big) \big(\frac{1}{S}\sum_{\mu=1}^S n_{j\mu}\big)}
\end{equation}
with $n_{i\mu}$ the abundance of ASV $\mu$ in station $i$, and $S$ the total number of ASVs in stations $i$ and $j$. To get the empirical correlation length, we fit the relation $\ln(g_{ij}) = -\frac{1}{\Gamma}d_{ij} + b$ from Eq. (\ref{eq:Correlation}) for each subset of the data in R using the package \texttt{tidyverse}~\cite{Wickham2019_tidyverse}.

\subsubsection*{GRUMP: Spatial Taylor's law and SAD}
For the analysis of the spatial Taylor's law, we use the same subset of filtered samples as the one used in the calculation of the spatial correlation length. The samples are ordered according to their geographical locations. The algorithm behind the spatial Taylor's law combines an increasing number of samples and calculates the mean and variance of ASV abundances in the combined samples. Assume that we have twelve samples ordered according to their geographical location. First, we calculate the mean and variance of the ASV abundance in the single samples. Next, we group the samples into pairs of two and recalculate the mean and variance for the six combined samples. Next, we do the same for the four triplets of samples, and so on until we get to the mean and variance of the ASV abundances in the whole transect combining all twelve samples. Using the average geographical distance between the samples, we estimate the spatial scale $L$ for each grouping of the samples, which then resulted in the empirical $Var(n_L)$. We fit the expression in Eq.~\ref{eq:results13} to this empirical result and then obtained an independent estimate for the correlation length $\Gamma$ and the Taylor's law exponent $\alpha$. Finally, we fit the GIG distribution to the ASV abundances in all samples of each data subset using the likelihood method with the minimization function of \texttt{SciPy} \cite{Virtanen2020_scipy}. In doing so, we obtain an estimate of $\lambda$. Subsequently, we fit the same set of samples using a lognormal distribution and compute the corresponding $\Delta_{\text{BIC}}$ values to compare the two distributions.

\subsubsection*{Tara Chlorophyll: patchiness}\label{SM:patchiness_obsevation}
The spatial correlation function for chlorophyll concentration is estimated by means of the Auto Correlation Function using the Python package \texttt{statsmodels}.  The covariance between two samples after a number $k$ of spatial lags is defined as follows:
\begin{equation}\label{eq:SM_ACF_pack}
    \text{ACF}(k)=\Omega^{-1}\sum_{i=1}^{\Omega-k}(n(i)-\bar{n})(n(i+k)-\bar{n})
\end{equation}
where $\Omega$ is the total number of points and $\bar{n}$ is the empirical average of $n$ on this sample. In our case, one lag corresponds to $200$~meters. In order to estimate Eq.~\ref{eq:SM_ACF_pack}, the code evaluates the Inverse Fourier Transformation of the power spectrum~\cite{william2007_numerical}. 
Using Eq.~\ref{eq:Correlation}, $\Gamma$ is estimated by fitting $\log\left(\text{ACF}(k)\right)=a+b\,k$ and taking $\Gamma=-b^{-1}$.

The patchiness $p_\textrm{obs}$ is estimated for each transect taking, for each $L$, the average value of $Var(n|L)$. Then we fit the power-law function $\log\left(Var(n|L)\right)=h+p_{\textrm{obs}}\log(L)$, where $h$ is a free variable. Similarly, the model correlation length $\Gamma_{\textrm{model}}$ is measured by fitting Eq.~\ref{eq:results14} in the log-log scale. To quantify the goodness of fits, we use the BIC. Specifically, we define $\Delta_{\textrm{BIC}}=\textrm{BIC}(\textrm{power-law})-\textrm{BIC}(\textrm{model})$. Naming $f$ a function of $L$, we use the following $\chi^2$ metric:
\begin{equation}
    \chi^2 = \sum_i^\Omega \Big(\log(Var(n|L_i))-f(L_i)\Big)^2
\end{equation}
where $\Omega$ is the number of fitted points. Secondly, assuming normality of the residuals on logarithmic scale, we have~\cite{Casella2024_statistical}:
\begin{equation}
    -2\log(\mathcal{L}) \sim \chi^2
\end{equation}
Therefore, $\text{BIC}$ can be written as:
\begin{equation}
    \text{BIC} \sim  \chi^2
\end{equation}
since the two models have the same number of parameters. Therefore, we arrive at the following expression:
\begin{equation}
    \Delta_{\text{BIC}} \sim \chi^2(\text{power-law}) - \chi^2(\text{model})
\end{equation}
We measure $\Delta_{\textrm{BIC}}$ for each transect: when $\Delta_{\textrm{BIC}}>0$, the saturation regime of $Var(n|L)$ is evident and Eq.~\ref{eq:results14} captures the patchiness of the systems; otherwise, the scaling of the variance is better described by the empirical power-law relation.

\begin{figure}
    \centering
    \includegraphics[width=0.78\linewidth]{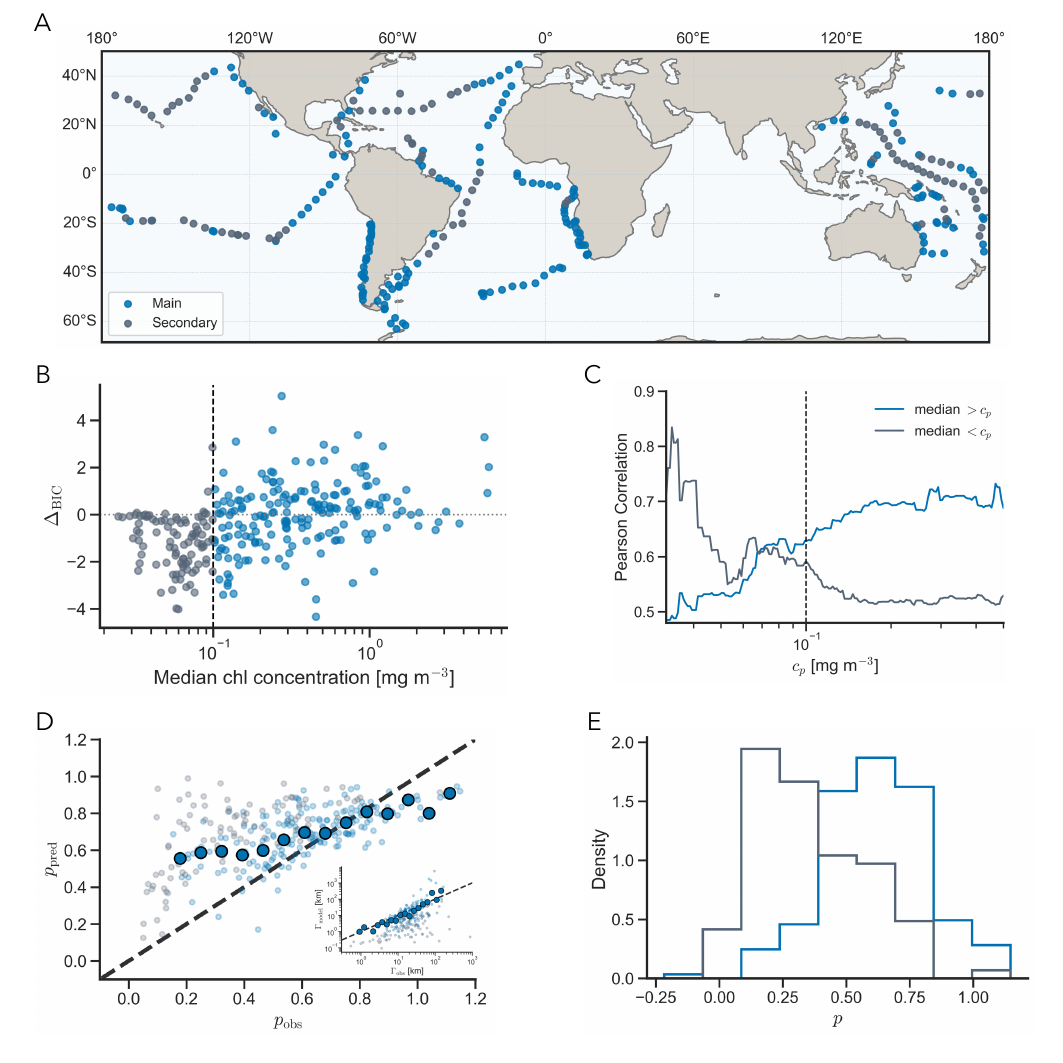}
    \caption{\textbf{Patchiness analysis.} (\textbf{A}): Geographic distribution of samples in upwelling regions with chlorophyll concentrations greater than 0.1~mg/$\textrm{m}^3$ (blue) and oligotrophic regions with concentrations less than the threshold. (\textbf{B}): $\Delta_{\textrm{BIC}}$ as a function of the median chlorophyll concentration for all transects. We note that there is a clear preference for the power-law (negative $\Delta_{\textrm{BIC}}$) among samples in the secondary cluster, while $\Delta_{\textrm{BIC}}$ appears to be equally distributed between positive and negative values in the main cluster. (\textbf{C}): Analysis of the Pearson correlation between $p_{\textrm{obs}}$ and $\Gamma_{\textrm{obs}}$ for different threshold values. For thresholds larger than $c_p=0.1$~mg/$\textrm{m}^3$, we have lower correlations for the secondary cluster, while they increase significantly for the main one. (\textbf{D}): We compare $p_\textrm{pred}$ (derived using $\Gamma_\textrm{obs}$) and $p_\textrm{obs}$. The large blue points represent the median values of $p_\textrm{pred}$ of the main cluster. In the inset, we show a comparison between the correlation length obtained by fitting Eq.\ref{eq:results14} and the one derived from the autocorrelation. Also in this case, blue dots are the median values of $\Gamma_{\textrm{model}}$ for the main cluster. (\textbf{E}): Histograms of patchiness exponents ($p_{\textrm{obs}}$) for the two clusters. For the secondary cluster, the distribution is peaked around lower values compared to the main cluster.}
    \label{fig:SM_patch}
\end{figure}

Our analysis reveals a clear relation between $\Delta_{\textrm{BIC}}$ and local chlorophyll abundance, estimated as the median concentration along the transect (panel~B of Fig.~\ref{fig:SM_patch}). In particular, we find a range of median concentration values characterized by $\Delta_{\textrm{BIC}}<0$ and low concentrations, and another where $\Delta_{\textrm{BIC}}$ is independent of total abundance. We also observe that samples with low chlorophyll concentrations are also those in which the Pearson correlation between $\Gamma$ and $p$ is lowest (panel~C). This suggests that there is a systematic difference in the system properties as a function of chlorophyll concentration. After exploring various threshold values $c_p$, we identify $c_p = 0.1~$ $\textrm{mg}/\textrm{m}^3$ as the critical value: for samples with higher concentrations, there is a strong correlation between $\Gamma$ and $p$ as expected (see also~\cite{mahadevan2002_biogeochemical}), whereas these variables show weaker correlations at lower concentration levels. We delineate the \emph{main} cluster as the group of samples with concentrations larger than $c_p$, while the \emph{secondary} cluster contains all the other samples. With these definitions, the main cluster includes approximately $66\%$ of the transects.  In panel~A of Fig.~\ref{fig:SM_patch}, we see the geographic distribution of the two groups of samples; in detail, the main cluster predominantly corresponds to upwelling regions, while the secondary samples are found in oligotrophic regions. In panel~D, we demonstrate that, for the main cluster, the correspondence between theoretical predictions and observational data is improved compared to the full dataset. In detail, the fit of $\Gamma_{model}$ using Eq.~\ref{eq:results14} is in perfect agreement with the correlation length estimated using the autocorrelation function ($\Gamma_{obs}$), and the theoretical prediction of $p$ better fits the empirical one measured from a power-law.  Therefore, our study suggests that, in upwelling regions, total abundance dynamics are consistent with a stochastic description coupled with a diffusive term, whereas in oligotrophic regions, where the biological signal is weaker, we hypothesize that the contribution of other factors, such as local heterogeneities of the biological rates, play a predominant role in defining the chlorophyll patchiness. Indeed, the secondary-cluster samples are also characterized by lower values of $p$ (panel~E), indicating that local variability dominates over large-scale fluctuations.

\subsubsection*{Tara Chlorophyll: total abundance distribution}\label{SM:Chl_tad}
To study the distribution of chlorophyll, we analyze all 1,000 samples collected along each transect. Therefore, the resulting distributions correspond to spatiotemporal series. Our analysis reveals that the majority of the samples (83\%) are better described by a log-normal distribution rather than by a GIG, according to~\cite{Campbell1995_lognormal}.

\subsubsection*{Tara Chlorophyll: $p$ as a function of the response time}\label{SM:patch}
\begin{figure}
    \centering
    \includegraphics[width=1\linewidth]{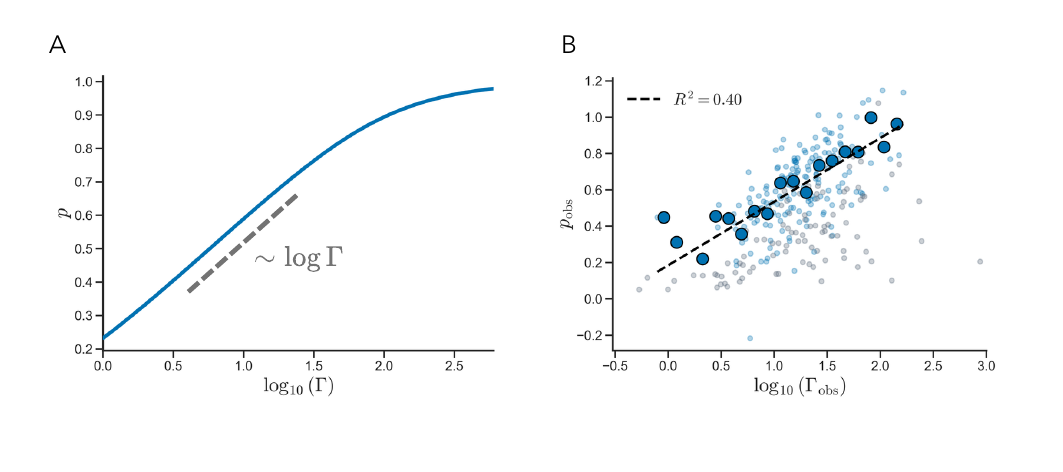}
    \caption{\textbf{Functional relation between patchiness and $\Gamma$.} (\textbf{A}): Theoretical functional relation between the patchiness slope and the correlation length given by Eq.~\ref{eq:SM_response}. We set $L_{\textrm{min}}=1$ and $L_{\textrm{max}}=300$. (\textbf{B}): Vertical axis represents the the observed patchiness exponent,  while the horizontal axis is the logarithm of the correlation length measured from the autocorrelation function. The large blue dots are the median values of $p_{\textrm{obs}}$ measured for each sample (small dots). The dashed line represents the linear fit derived considering the data points belonging to the main cluster.}
    \label{fig:SM_response}
\end{figure}

We conclude our study of the spatial heterogeneity of chlorophyll by testing the scaling of the patchiness exponent $p$ with respect to the plankton response time $\tau$. Indeed, several works (e.g.~\cite{mahadevan2002_biogeochemical, Bracco2009_horizontal}) have highlighted a relationship between $p$ and  $\tau$. In detail, in~\cite{mahadevan2002_biogeochemical} the authors have found a simple relation: $p\propto \log{\tau}$. Using the Tara dataset and our analytical results, we reconfirm this behavior. Recalling Eq.~\ref{eq:results14}, we adopt $Var(n)=1$ to simplify the calculations and set the minimum observable spatial scale to $L_{\textrm{min}}=1$. We denote the maximum spatial scale by $L_{\textrm{max}}$ and assume $1\ll\Gamma\ll L_{\textrm{max}}$, where $\Gamma$ is the correlation length. Then, using Eq.~\ref{eq:results14} we have $Var(n|L_{\textrm{max}})\approx 1$ and $Var(n|1)\approx (3 \Gamma)^{-1}$. Under these assumptions, the model patchiness $p$ can be estimated as follows:
\begin{equation}\label{eq:SM_response}
    p(\Gamma)\approx \frac{\log Var(n|L_{\max})-\log Var(n|L_{\min})}{\log(L_{\max}/L_{\min})}\approx\frac{\log{\left(3\Gamma\right)}}{\log{\left(L_{\textrm{max}}\right)}}
\end{equation}
This expression is the slope of the line that passes through $Var(n|L_{\textrm{max}})$ and $Var(n|1)$ as a function of the window size on log-log scale. Therefore, Eq.~\ref{eq:SM_response} provides an analytical relationship between $p$ and $\Gamma$: as shown in panel~A of Fig.~\ref{fig:SM_response}, for $\Gamma\ll L_{\textrm{max}}$ we see that $p$ is proportional to $\log(\Gamma)$. Since by definition $\log(\Gamma)\propto \log(\tau) $, this proves the expected linear relation with the logarithm of the response time. This trend is also recognizable in the main cluster of Tara Pacific-Tara Micriobiome dataset, where we find a linear relationship ($R^2=0.4$) between  $p_{\textrm{obs}}$ and $\log(\Gamma_{\textrm{obs}})$ (panel~B of Fig.~\ref{fig:SM_response}).  We note that for the secondary cluster, the correlation between $p$ and $\Gamma$ is less evident, further confirming that, for oligotrophic samples, spatial variability cannot be described solely in terms of an environmental noise coupled with diffusion.




\clearpage 

\end{document}